\documentclass[utf8]{frontiersSCNS}
\usepackage{url,hyperref,lineno,microtype,subcaption,bm}
\usepackage[onehalfspacing]{setspace}
\usepackage[normalem]{ulem}
\DeclareMathOperator*{\argmin}{argmin}
\def\keyFont{\fontsize{8}{11}\helveticabold}
\def\firstAuthorLast{Thiem {et~al.}}
\def\Authors{Thomas N.~Thiem\,$^{1}$, Mahdi Kooshkbaghi\,$^{2}$, Tom Bertalan\,$^{3}$, Carlo R. Laing\,$^{4}$, and Ioannis G.~Kevrekidis\,$^{5,*}$}
\def\Address{$^{1}$Department of Chemical and Biological Engineering, 
Princeton University, USA \\
$^{2}$Program in Applied and Computational Mathematics,
Princeton University, USA \\
$^{3}$Department of Mechanical Engineering,
Massachusetts Institute of Technology, USA \\
$^{4}$School of Natural and Computational Sciences,
Massey University, NZ \\
$^{5}$Department of Applied Mathematics and Statistics, 
Johns Hopkins University, USA
}
\def\corrAuthor{Ioannis G. Kevrekidis}
\def\corrEmail{yannisk@jhu.edu}

\begin{document}
\onecolumn
\firstpage{1}

\title[Emergent spaces for coupled oscillators]{Emergent spaces for coupled oscillators} 
\author[\firstAuthorLast]{\Authors}
\address{}
\correspondance{}
\extraAuth{}

\maketitle

\begin{abstract}
Systems of coupled dynamical units (e.g. oscillators or neurons) are known to exhibit complex, emergent behaviors that may be simplified through coarse-graining: a process in which one discovers coarse variables and derives equations for their evolution. Such coarse-graining procedures often require extensive experience and/or a deep understanding of the system dynamics. In this paper we present a systematic, data-driven approach to discovering ``bespoke" coarse variables based on manifold learning algorithms. We illustrate this methodology with the classic Kuramoto phase oscillator model, and demonstrate how our manifold learning technique can successfully identify a coarse variable that is one-to-one with the established Kuramoto order parameter. We then introduce an extension of our coarse-graining methodology which enables us to learn evolution equations for the discovered coarse variables via an artificial neural network architecture templated on numerical time integrators (initial value solvers). This approach allows us to learn accurate approximations of time derivatives of state variables from sparse flow data, and hence discover useful approximate differential equation descriptions of their dynamic behavior. We demonstrate this capability by learning ODEs that agree with the known analytical expression for the Kuramoto order parameter dynamics at the continuum limit. We then show how this approach can also be used to learn the dynamics of coarse variables discovered through our manifold learning methodology. In both of these examples, we compare the results of our neural network based method to typical finite differences complemented with geometric harmonics. Finally, we present a series of computational examples illustrating how a variation of our manifold learning methodology can be used to discover sets of ``effective'' parameters, reduced parameter combinations, for multi-parameter models with complex coupling. We conclude with a discussion of possible extensions of this approach, including the possibility of obtaining data-driven effective partial differential equations for coarse-grained neuronal network behavior, as illustrated by the synchronization dynamics of Hodgkin-Huxley type neurons with a Chung-Lu network. Thus, we build an integrated suite of tools for obtaining data-driven coarse variables, data-driven effective parameters, and data-driven coarse-grained equations from detailed observations of networks of oscillators.
\tiny
\keyFont{\section{Keywords:} Diffusion Maps, Manifold Learning, Geometric Harmonics, Neural Networks, Kuramoto Oscillators, Coupled Systems, Complex Networks}
\end{abstract}

\section*{Introduction}
\label{sec:Introduction}
We study coupled systems comprised of many individual units that are able to interact to produce new, often complex, emergent types of dynamical behavior. The units themselves (motivated by the modeling of large, complex neuronal networks) may be simple phase oscillators, or may be much more sophisticated, with heterogeneities and parameter dependence contributing to the emerging behavior complexity. Each unit is typically described by a system of ordinary differential equations (ODEs) \eqref{eq:coupled_system_ex_ODEs}, where $\mathbf{f}$ is a function of the system state $\textbf{x}$, time $t$, and the system parameters \textbf{p},
\begin{equation}
\frac{d\textbf{x}}{dt}=\mathbf{f}(\textbf{x}, t; \textbf{p}).
\label{eq:coupled_system_ex_ODEs}
\end{equation}
Examples of such systems from across the scientific domains range from coupled reactor networks \citep{mankin1986dynamics} and molecular dynamics simulations \citep{Haile:1997:MDS:531139}, to the modeling of synchronization and swarming among oscillators \citep{o2017oscillators}, and even to the oscillations of the millennium bridge \citep{strogatz2005theoretical}. For large coupled oscillator ensembles, numerically evolving the system state can be computationally expensive; yet several such systems of practical interest feature an underlying structure that allows them to be described through a small collection of \textit{coarse variables} or \textit{order parameters} whose dynamic evolution may be described in simpler, reduced terms. It is also observed that the dynamics of these coarse variables may only practically depend on a reduced set of parameters, the ``effective'' parameters of the system, which themselves are specific, possibly nonlinear, combinations of the original, detailed system parameters. In general, uncovering such coarse, descriptive variables and parameters, and approximating their effective dynamic evolution is a significant undertaking, typically requiring extensive experience, deep insight, and painstaking theoretical and/or computational effort.

Here we present and illustrate an alternative, automated, data-driven approach to finding emergent coarse variables and modeling their dynamical behavior. As part of our methodology we make use of manifold learning techniques, specifically diffusion maps (DMAPs) \citep{coifman2006diffusion}, to systematically infer tailor-made descriptive variables directly from detailed dynamical data itself, without any prior knowledge of the underlying models/equations. This stands in stark contrast to established, intuition-based or equation-based coarse-graining methods, which rely on a combination of system knowledge, experience and strongly model-dependent mathematical techniques to invent or derive the coarse variables/equations.

We begin our presentation with a brief outline of our key manifold learning tool, the diffusion maps technique, followed by an abridged introduction to the associated geometric harmonics function extension method \citep{coifman2006geometric}. Following these introductions, we present a demonstration of our data-driven methodology for coarse variable identification. After this, we consider the time dependent behavior of coarse variables, and show how their corresponding coarse dynamic evolution equations can be learned from time series of dynamical observation data by virtue of an artificial neural network architecture templated on numerical time integration schemes. We compare this approach to typical finite difference based methods complemented with geometric harmonics. Finally, we illustrate how manifold learning techniques can be used to find ``effective,'' reduced sets of parameters in multi-parameter coupled systems.

Motivated by the intended applications of our methodology to neurological systems, we choose to illustrate our techniques with one of the simplest neurobiologically salient models available, the classic Kuramoto coupled phase oscillator model \citep{kuramoto1975self, kuramoto1984}. The Kuramoto model has become a popular choice to study synchronization \citep{schmidt2014dynamics} and network topology in neuroscience \citep{rodrigues2016kuramoto}. At first glance the scalar phase dynamics of the model may seem to be insufficient or highly restrictive. However, the phase reduction approach has become a standard technique in computational neuroscience \citep{ermentrout1986parabolic, ermentrout1990oscillator, brown2004phase, tass2007phase, guckenheimer2013nonlinear} providing a link between computational models of neurons and models of weakly coupled phase oscillators. Indeed, a classical example of this approach is in the characterization of a simple (class I) spiking neuron as a one dimensional phase oscillator \citep{ermentrout1986parabolic, ermentrout1996type}. Recently, phase based measures of synchrony have become a typical feature in the characterization of large-scale experimental neuroscience signals \citep{mg1998weule, varela2001brainweb, breakspear2002nonlinear, stam2007phase, penny2009dynamic}. Advances in the field of neural mass models have furnished standard approcahes to describe the interaction between excitatory and inhibitory neurons, such as the Wilson-Cowan model \citep{wilson1973mathematical}. Weakly-coupled Wilson-Cowan oscillators have subsequently been shown to exhibit similar interaction dynamics to weakly-coupled Kuramoto oscillators \citep{izhikevich1997weakly}. The main deficiency of the Kuramoto model is the lack of a spatial embedding for the oscillators; however numerous modifications of the interaction term of the model have been proposed to circumvent this difficulty, such as time delays, distance-dependent transmission delays, finite-support wavelet-like spatial kernels, and second order phase interaction curves \citep{breakspear2010generative}. These modifications yield rich spatiotemporal behaviors, such as traveling rolls and concentric rings \citep{jeong2002time}, which are similar to those observed {\em in vivo} \citep{freemann1975mass, prechtl1997visual, lam2000odors, du2005encoding, rubino2006propagating}. Furthermore, these modified models have been used to understand dynamical behavior on cortical-like sheets \citep{honey2007network, deco2009key}, simulate the BOLD signal \citep{cabral2011role}, and study hypersynchronous neural activity \citep{schmidt2014dynamics}.

We make use of the classic Kuramoto coupled phase oscillator model and its variations throughout this paper as prototypical examples to showcase our methodology. In our concluding discussion and future work section we also mention the possibility of discovering effective emergent \textit{partial differential equation} descriptions of heterogeneous network dynamics, illustrated through the synchronization of Hodgkin-Huxley type neurons on a Chung-Lu type network.
\section*{Diffusion Maps}
\label{sec:DiffusionMaps}
Introduced by Coifman and Lafon \citep{coifman2006diffusion}, the diffusion maps (DMAPs) method and associated algorithms form part of a class of dimensionality reduction approaches, techniques which are used to find the intrinsic dimensionality of high dimensional data. Specifically, the diffusion map (DMAP) algorithm is a manifold learning technique that seeks to address the problem of parametrizing $d$-dimensional manifolds embedded in $\mathbb{R}^n$ based on data, with $d < n$. It accomplishes this by constructing a (discretized) Laplace operator on the data, such that the operator's eigenfunctions define embedding coordinates for the manifold. An algorithm for numerically constructing this operator is provided by Coifman and Lafon \citep{coifman2006diffusion}.

At the heart of the Laplace operator construction lies the kernel, $k$. The kernel describes the affinity/similarity between data points and serves to define the local geometry of the underlying manifold. This information is frequently stored in a kernel matrix $K$, where $K_{ij}$ is the kernel evaluated on data points $i$ and $j$. As the specific parametrization intimately depends on the data geometry, it is essential to choose a kernel that captures the relevant properties of the data. An example of a typical Gaussian kernel with the Euclidean distance is shown in \eqref{eq:dmap_kernel},
\begin{equation}
K_{ij} = k(x_{i}, x_{j}) = \exp\left(- \frac {\Vert x_i-x_j\Vert_2^2}{\epsilon^2}\right)\quad i,j=1,\dots,n,
\label{eq:dmap_kernel}
\end{equation}
where $\epsilon$ is a tunable kernel bandwidth parameter and $\Vert \cdot\Vert_2$ is the Euclidean norm. By selecting different kernel bandwidth parameters it is possible to examine features of the data geometry at different length scales. Several heuristics are available to select the value of the kernel bandwidth parameter \citep{dsilva2018parsimonious}; here we select the median of the pairwise distances between the data points $x_{i}$ as a starting point for our $\epsilon$ tuning.

After defining a kernel, the Laplace operator can be approximated as follows. First, one forms the diagonal matrix $D$ with $D_{ii}=\sum_{j}K_{ij}$, which is then used to construct the matrix $\widetilde{K}=D^{-\alpha}KD^{-\alpha}$, where $\alpha$ is an algorithm parameter; the choice of $\alpha$ determines the form of the operator that is approximated, which in turn dictates the influence of the sampling density of the data on the parametrization of the underlying manifold. Common choices are: (a) $\alpha=0$, the normalized graph Laplacian (the one most influenced by the sampling density and useful for uniformly sampled manifolds), (b) $\alpha=0.5$, Fokker-Planck diffusion, and (c) $\alpha=1$, the Laplace-Beltrami operator (which removes the influence of the sampling density). Next, one constructs the diagonal matrix $\widetilde{D}$ with $\widetilde{D}_{ii}=\sum_{j}\widetilde{K}_{ij}$ and uses this to form the matrix $A=\widetilde{D}^{-1}\widetilde{K}$. The eigenvectors of $A$ provide new \textit{intrinsic} coordinates for the data points on the manifold.

A challenge of using the eigenfunctions of a Laplace operator to parametrize the manifold on which the data lie is the existence of ``repeated" coordinates, that is, coordinates which parametrize an already discovered direction along the manifold, referred to as \textit{harmonics} \citep{dsilva2018parsimonious}. These harmonics provide no additional dimensional information, and should be ignored in an effort to discover a minimal embedding. A systematic computational method has been developed for automatically identifying such harmonics via a local linear regression on the eigenfunctions \citep{dsilva2018parsimonious}.

Before the local linear regression method can be applied, the eigenfunctions of the Laplace operator must be sorted by their corresponding eigenvalues, all of which are real, from greatest to smallest. The main idea behind the local linear regression method is that the repeated eigenfunctions, the harmonics, can be represented as functions of the fundamental eigenfunctions, which correspond to unique directions. 
The local linear regression method checks for this functional relationship by computing a local linear fit \eqref{eq:local_linear_regression} of a given eigenfunction $\phi_{k} \in \mathbb{R}^{d}$ (where $d$ is the number of data points) as a function of the previous eigenfunctions $\mathbf{\Phi}_{k-1}=[\phi_{1} \cdots \phi_{k-1}]^T$(those previously discovered) for each data index $i$.
\begin{equation}
\phi_{k}(i)\approx \alpha_{k}(i)+\beta_{k}^{T}(i)\mathbf{\Phi}_{k-1}(i)
\label{eq:local_linear_regression}
\end{equation}
The coefficients of the fit, $\alpha_{k}(i) \in \mathbb{R}$ and $\beta_{k}(i) \in \mathbb{R}^{k-1}$, vary over the domain (indexed by the data points $i$) and are found by solving the following optimization problem \eqref{eq:local_linear_optimization}.
\begin{equation}
\alpha_{k}(i), \beta_{k}(i) = \argmin_{\alpha, \beta}\sum_{j \neq i} K(\mathbf{\Phi}_{k-1}(i),\mathbf{\Phi}_{k-1}(j))(\phi_{k}(j)-(\alpha+\beta^{T}\mathbf{\Phi}_{k-1}(j)))^{2}
\label{eq:local_linear_optimization}
\end{equation}
The weighting kernel $K$ for the local linear regression \eqref{eq:local_linear_optimization} is responsible for the domain dependence of the coefficients and is typically Gaussian \citep{dsilva2018parsimonious}, such as in \eqref{eq:local_linear_kernel}. For example
\begin{equation}
K(\mathbf{\Phi}_{k-1}(i),\mathbf{\Phi}_{k-1}(j))=\exp\left(- \frac {\Vert \Phi_{k-1}(i)-\Phi_{k-1}(j)\Vert_2^2}{\epsilon^{2}_{reg}}\right),
\label{eq:local_linear_kernel}
\end{equation}
where $\epsilon_{reg}$ is a tunable scale for the kernel of the regression. Choosing $\epsilon_{reg}$ to be one third of the median of the pairwise distances between the eigenfunctions $\mathbf{\Phi}_{k-1}(i)$ is recommended as a starting point \citep{dsilva2018parsimonious}.

The quality of the fit is assessed by the normalized leave-one-out cross-validation error \eqref{eq:local_linear_residual} or the residual $r_{k}$ of the fit. Values of the residual near zero indicate an accurate approximation of $\phi_{k}$ from $\mathbf{\Phi}_{k-1}$, indicative of a harmonic eigenfunction, while values near one correspond to a poor fit, and are therefore suggestive of a significant, informative eigenfunction corresponding to a new direction in the data. Thus, by computing residual values for a collection of the computed eigenfunctions, it is possible to identify the fundamentals and the harmonics in a systematic way,
\begin{equation}
r_{k}=\sqrt{\frac{\sum_{i=1}^{n}(\phi_{k}(i)-(\alpha_{k}(i)+\beta_{k}(i)^T\mathbf{\Phi}_{k-1}(i)))^2}{\sum_{i=1}^{n}(\phi_{k}(i))^2}}.
\label{eq:local_linear_residual}
\end{equation}
Throughout this paper we employ the DMAPs algorithm to identify and parametrize manifolds as well as the local linear regression method to identify the significant, fundamental (non-harmonic) eigenfunctions. In all cases, we use $\alpha=1$ to remove the influence of the sampling density of the data, and the Gaussian kernel with the Euclidean distance defined in \eqref{eq:dmap_kernel} as our similarity measure. For the local linear regression, we utilize a Gaussian kernel \eqref{eq:local_linear_kernel} and select $\epsilon_{reg}$ as one third of the median of the pairwise distances between the $\mathbf{\Phi}_{k-1}$ as our regression kernel scale, in accordance with \citep{dsilva2018parsimonious}.
\section*{Geometric Harmonics}
\label{sec:GeoHarmonics}
Consider a set $X$ sampled from a manifold $\bar{X} \subset \mathbb{R}^N$ with finite measure $\mu(X)<\infty$ for some measure $\mu$. Let us assume that we have a real valued function $F:X\rightarrow\mathbb{R}$ defined on $X$. Geometric harmonics is a tool for extending $F$ to points on the original manifold $\bar{X}$, namely, it provides a way to find a new function $f:\bar{X}\rightarrow \mathbb{R}$, which can be considered as an extension of $F$ to $\bar{X}$ \citep{coifman2006geometric}.

The first step of computing the geometric harmonics extension is to define a kernel $k: \bar{X} \times \bar{X} \to \mathbb{R}$, where $k$ need not be the same as in \eqref{eq:dmap_kernel}. The kernel must be symmetric and bounded on the data set. An example of such a kernel is the Gaussian kernel with the Euclidean distance defined in \eqref{eq:dmap_kernel}. The main features of this kernel that are required are: (a) the ability to easily evaluate it on out-of-sample data points, and (b) that its eigenfunctions form a basis of a function space. By utilizing these features it is possible to represent a function in terms of the eigenfunctions of this kernel (as approximated by the data) and then extend it to out-of-sample data points.

Consider $n$ data points at which we know the values of the function of interest, $F$. We want to approximate $F$ at some $m$ other, new points by $d$ geometric harmonics. We can construct a matrix $K\in\mathbb{R}^{m\times n}$ with entries $K_{ij}=k(x_i,x_j)$, where $x_i$ is an unknown point, $x_j$ is a sampled point with known value of $F(x_{j})$, and $k$ is the kernel. The $l$-th geometric harmonic is defined by
\begin{equation}
    \Phi_i^{(l)}=\Phi^{(l)}(x_i) = \sum_{j=1}^{n} K_{ij}\psi_j^{(l)}\lambda_l^{-1},
\end{equation}
where $\psi^{(l)}$ is the $l$-th column of the matrix $\Psi \in \mathbb{R}^{n\times d}$, the first $d$ eigenfunctions of the kernel matrix, and $\lambda_l \in \Lambda=diag(\lambda_1, \dots, \lambda_d)\in \mathbb{R}^{d \times d}$ is the corresponding eigenvalue. The extension $f$ of the known function $F$ to the set of new points is then computed as \citep{coifman2006geometric}
\begin{equation}
    f = K\Psi \Lambda^{-1} \Psi^T F.
\end{equation}
Any observable of the data (e.g. the phase or the amplitude or any state variable for each particular oscillator in our network) is a function on the low-dimensional manifold, and can therefore be extended out-of-sample through geometric harmonics. This provides us a systematic approach for translating back-and-forth between physical observations and coarse-grained ``effective" or ``latent space" descriptions of the emergent dynamics.
\section*{Identifying Coarse Variables (Order Parameters)}
\label{sec:Identifying Coarse Variables}
Here we illustrate how manifold learning techniques, in this case diffusion maps, can be used to identify coarse variables for coupled oscillator systems directly from time series data. We consider the simple Kuramoto model with sinusoidal coupling as a test case, to demonstrate how we can learn a variable that is equivalent to the typical Kuramoto order parameter, $R$.
\subsection*{The Kuramoto Model}
The Kuramoto model is a classical example of limit cycle oscillators that can exhibit synchronizing behavior. The basic version of the model consists of a number of heterogeneous oscillators ($i=1,\dots, N$), each with their own natural frequency $\omega_{i}$, selected from a distribution $g(\omega)$, that interact through a coupling term $f$. The coupling term is typically sinusoidal and can be expressed as a function of the difference in phases between oscillators $f(\theta_{i}-\theta_{j})$.

The strength and presence of coupling among the oscillators is expressed by the coupling matrix $A$. Each element of the matrix $A_{ij}\  (i,j = 1, \dots, N)$ is a pairwise strength quantifying the influence of oscillator $j$ on oscillator $i$. A general Kuramoto model with a sinusoidal coupling function can be written as \eqref{eq:general_kuramoto}
\begin{equation}
\frac{d\theta_i}{dt}=\omega_i+\frac{1}{N}\sum_{j=1}^N A_{ij} \sin(\theta_j-\theta_i) \quad \text{for}\ i=1,\dots, N.
\label{eq:general_kuramoto}
\end{equation}
Originally, Kuramoto considered a mean-field coupling approximation \citep{strogatz2000kuramoto}, $A_{ij}=K>0$ where $K$ is a global coupling constant, which results in the following simplified all-to-all coupled model
\begin{equation}
\frac{d\theta_i}{dt}=\omega_i+\frac{K}{N}\sum_{j=1}^N \sin(\theta_j-\theta_i) \quad \text{for}\ i=1,\dots, N.
\label{eq:simple_kuramoto}
\end{equation}

\begin{figure}[h]
\centering
\begin{tabular}{cc}
    \includegraphics[width=0.45\textwidth]{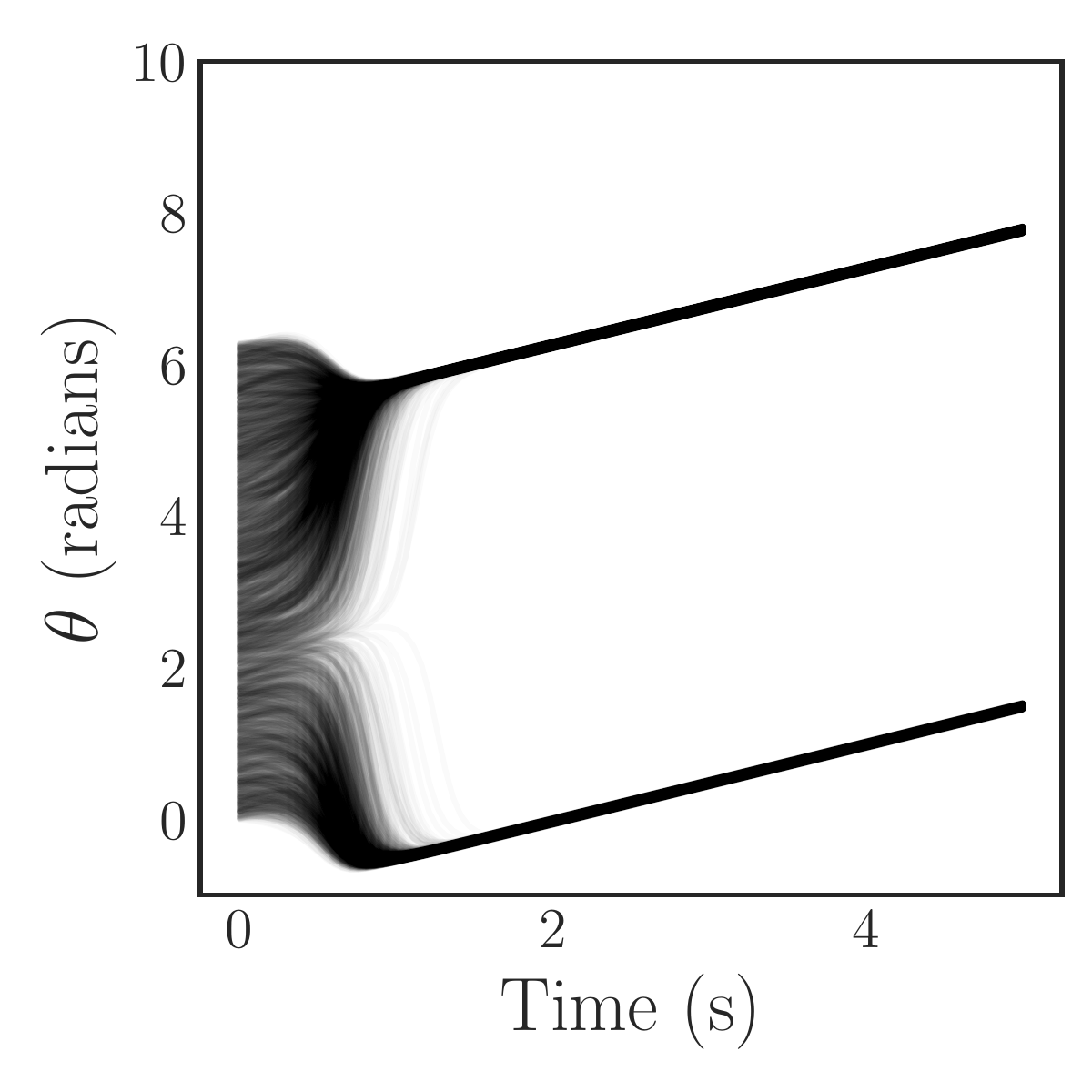}  &
    \includegraphics[width=0.45\textwidth]{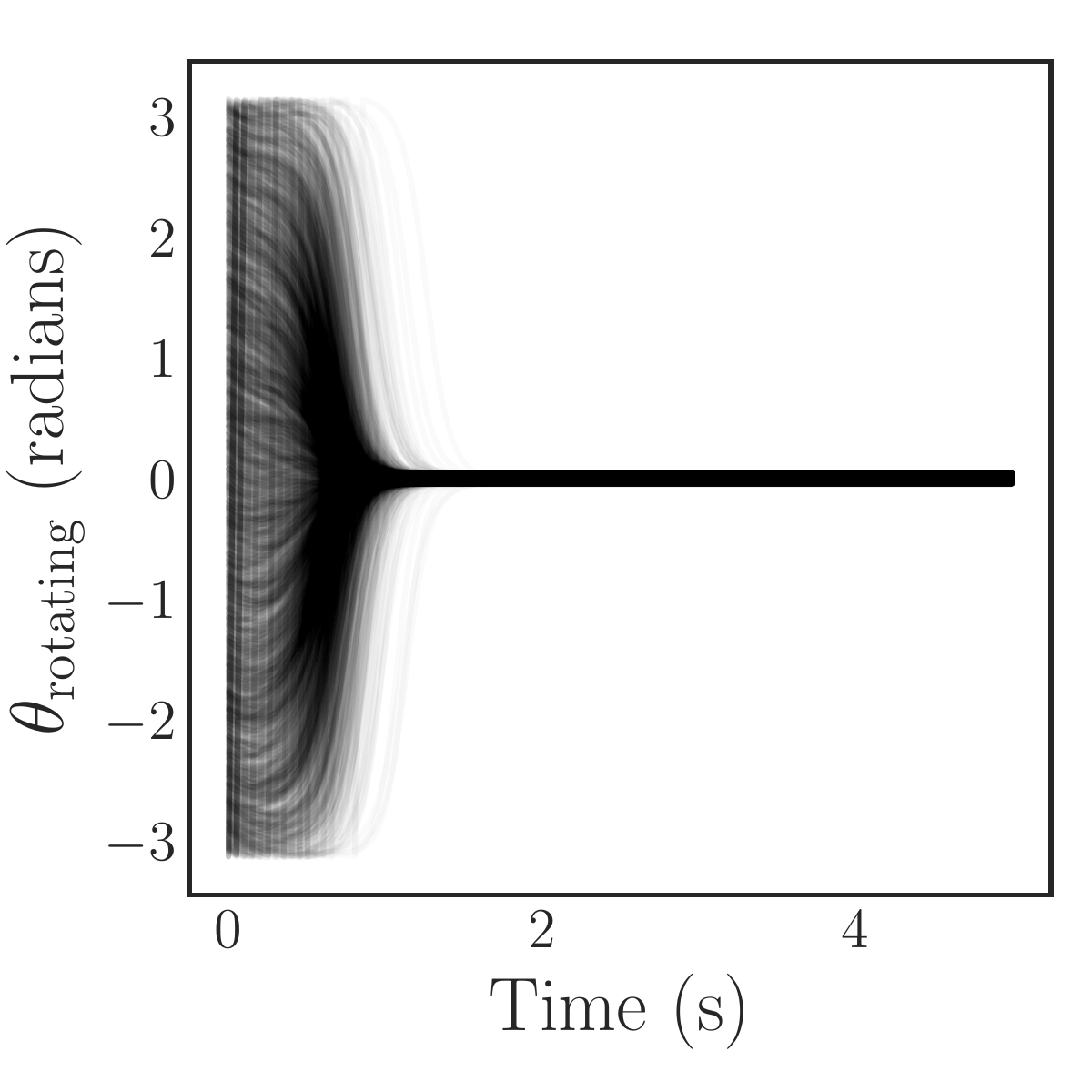} \\
    (a) & (b)
\end{tabular}
\caption{Time series of $N=4000$ Kuramoto oscillator phases with coupling that is strong enough to induce complete synchronization. The oscillators experience a short transient before synchronizing and converging to a steady state. This behavior is depicted for both a stationary frame (a), and a rotating frame (b).}
\label{fig:Simple_Kuramoto_time_series}
\end{figure}

The degree of phase synchronization of the oscillators can be expressed in terms of the complex-valued order parameter (a coarse variable) introduced by Kuramoto as follows \citep{kuramoto1984}
\begin{equation}
R(t)e^{i\psi(t)}=\frac{1}{N}\sum_{j=1}^N e^{i\theta_j(t)},
\end{equation}
where $0 \leq R(t) \leq 1$ is the time-dependent phase coherence, and $\psi(t)$ is the average phase. Values of the phase coherence near one correspond to phase synchronization of the oscillators, while values near zero imply disorder. An example of the typical synchronizing behavior of the Kuramoto model is illustrated for both a stationary reference frame \textbf{Fig.~\ref{fig:Simple_Kuramoto_time_series} (a)}, and a rotating reference frame \textbf{Fig.~\ref{fig:Simple_Kuramoto_time_series} (b)}. Note that in the rotating frame the frequency synchronization induces a steady state. 
In general, if the coupling is chosen such that the system exhibits complete synchronization (characterized by the lack of ``rogue" or unbound oscillators), then a steady state exists in a rotating reference frame. Typically, the rotating frame transformation is taken to be $\theta_i(t) \mapsto \theta_i(t)-\psi(t)$, in which $\psi(t)$ is the average phase. This transformation results in a system for which the new average phase is zero, yielding a real-valued order parameter. The order parameter in this rotating frame is given by
\begin{equation}
R(t)=\frac{1}{N}\sum_{j=1}^{N}e^{i(\theta_{j}(t)-\psi(t))}.
\end{equation}

Throughout the remainder of this paper we employ the simple all-to-all coupled Kuramoto model and its variations, some with more complicated couplings, to illustrate our methodology. We emphasize the synchronizing behavior of the Kuramoto model as it facilitates coarse-graining, and repeatedly make use of this property to simplify our calculations throughout the rest of the paper.
\subsection*{Order Parameter Identification}
For this section we consider the simple Kuramoto model with all-to-all mean-field coupling \eqref{eq:simple_kuramoto}. Our goal is to: (a) first identify a coarse variable from time series of phase data; and then (b) demonstrate that this discovered variable is equivalent to the established Kuramoto order parameter. Throughout, we consider a rotating frame in which only the magnitude of the order parameter $R$ varies in time. This in turn enables us to neglect the angle $\psi$, as $R$ suffices to capture the time-dependent behavior.

\begin{figure}[h]
    \centering
    \includegraphics[width=0.5\textwidth]{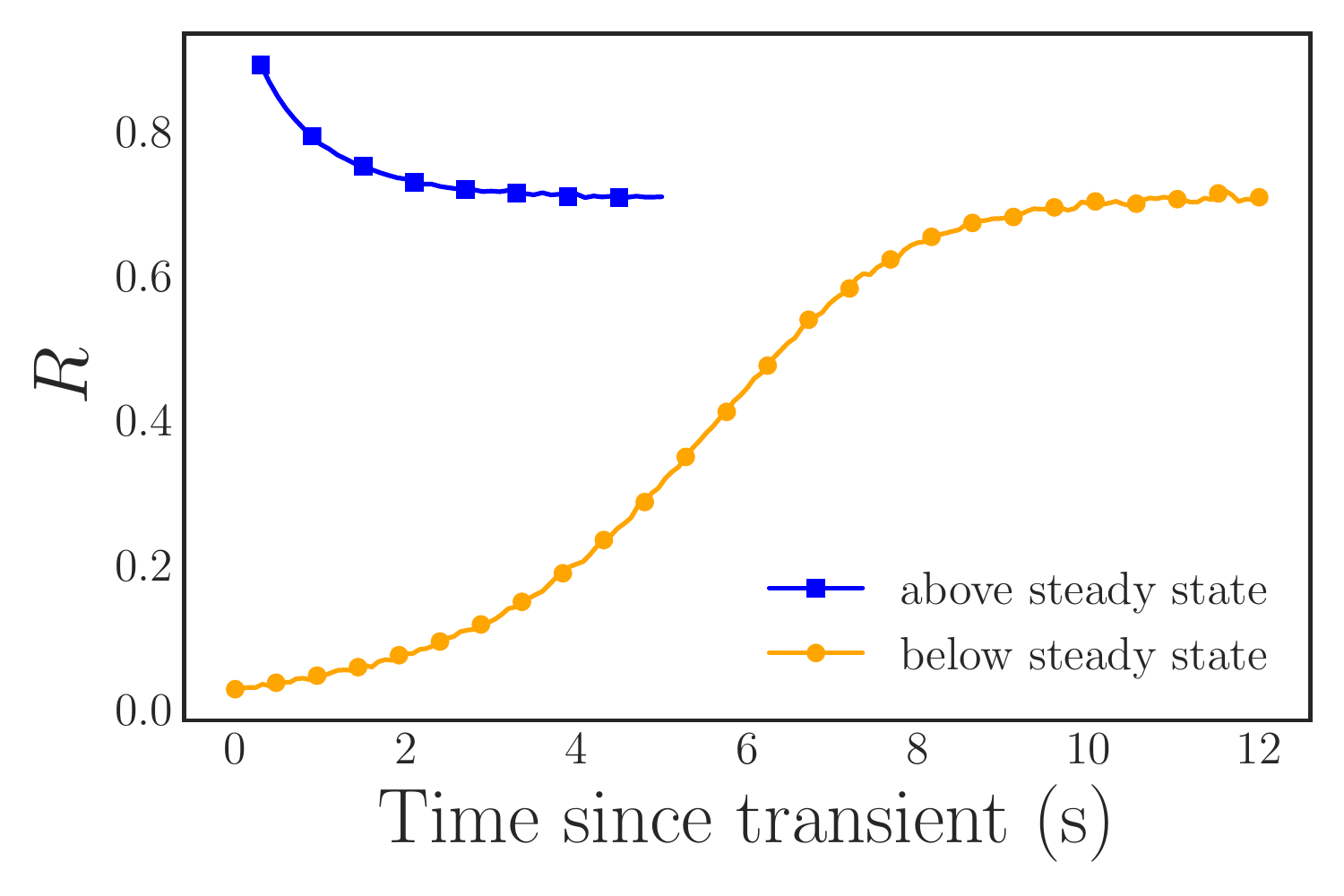}
    \caption{Plot of the magnitude $R$ of the Kuramoto order parameter for our two simulation trajectories after the rotating frame transformation: (blue) the trajectory initialized ``above" the steady state $R$ value, (orange) the trajectory initialized ``below" the steady state $R$ value. In both cases, the initial fast transients have already been discarded. The included markers highlight a subset of the data for clarity and are not representative of the entire data set. Note that the angle $\psi$ is time-independent in this rotating frame ($\psi(t)=0$).}
    \label{fig:op_trajectories}
\end{figure}

We begin our analysis by simulating 8000 Kuramoto oscillators ($N=8000$) with a coupling constant $K=2$, and frequencies drawn from a Cauchy distribution \eqref{eq:cauchy_distribution} with $\gamma=0.5$.
\begin{equation}
    g(\omega;\gamma)=\frac{\gamma/\pi}{\omega^2+\gamma^2}
    \label{eq:cauchy_distribution}
\end{equation}
In order to reduce the finite sample noise in $R$, we use inverse transform sampling, with equally spaced values over the interval $[0+\epsilon, 1-\epsilon]$ with $\epsilon \approx 2.5 \times 10^{-4}$, to generate our frequencies in a systematic and symmetric manner. We select $\epsilon$ to place a cap on the maximum absolute value of the frequencies in order to facilitate numerical simulation. For these parameter values, $R$ exhibits an attracting, stable steady state with $R_{\infty}\approx 0.71$. Thus, in order to sample the entire range of potential $R$-values, we select two different sets of initial conditions for our simulations, with one ``above" the steady state synchronization $R$ value, and the other ``below" it. For the first case, $R=1$, we set all of the initial oscillator phases to $\pi$ and for the second case, $R\approx 0$, we use equally spaced initial phases over $[0, 2\pi]$. In both cases, we integrate our system of oscillator ODEs with SciPy's Runge-Kutta integrator (\textit{solve\_ivp} with the RK45 integrator) with the absolute and relative tolerances set to $10^{-7}$ and $10^{-4}$, respectively. After discarding the initial transients, we transform the phase data into a rotating frame and then sample it at discrete, equidistant time steps to form our time series data, \textbf{Fig.~\ref{fig:op_trajectories}}.

Before applying the DMAPs algorithm to these time series of phase data to identify a coarse variable, we need to select a suitable kernel. It is crucial to select a kernel that will compare the relevant features of the data in order to produce a meaningful measure of the similarity between data points. In this case, it is important to choose a kernel that measures the degree of clustering or equivalently the phase synchronization of the oscillators \textit{in a way that is invariant to permutations of the oscillator identity}, as a relabeling of the oscillators should correspond to the same degree of synchronization.

\begin{figure}[h]
    \centering
    \includegraphics[width=0.75\textwidth]{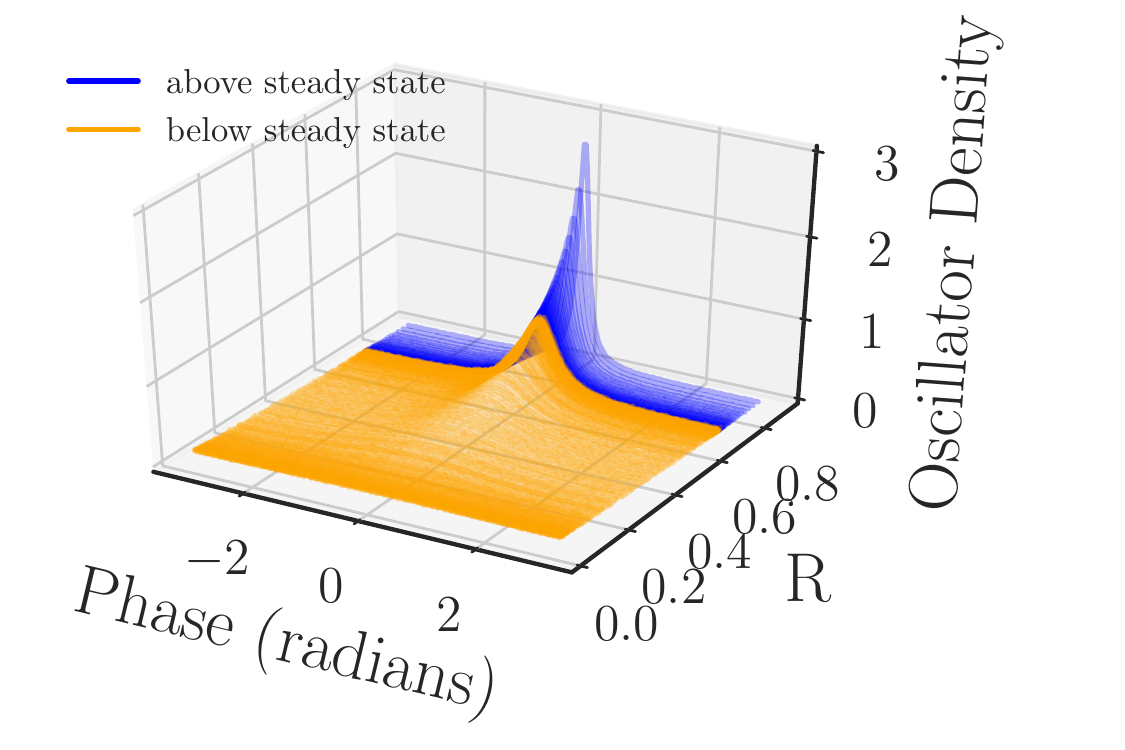}
    \caption{Normalized oscillator densities of the Kuramoto oscillator phases sampled at equal time intervals from the two simulation runs. The densities are colored by whether they originated from the simulation run that began above/below the steady state (blue/orange) $R$ value, and sorted by their corresponding value of $R$ (computed from their phases). All of the densities are computed from a histogram with 200 bins.}
    \label{fig:kura_density}
\end{figure}

Since the synchronization of the oscillators is related to how the oscillator phases group together, it is a natural choice to consider the phase density as a meaningful observable. The oscillator phase density captures phase clustering while being permutation independent, and should thus provide a meaningful similarity measure between system snapshots. Therefore, we first pre-process the phase data by computing the density of the oscillators over the interval $[-\pi, \pi]$ before defining the kernel. We approximate this density with a binning process (histogram) that uses 200 equally spaced bins over the interval $[-\pi, \pi]$. As part of the binning process we are careful to reduce the phases of the oscillators mod $2\pi$ to ensure that all of them are captured in the $[-\pi, \pi]$ interval, \textbf{Fig.~\ref{fig:kura_density}}.

\begin{figure}[h]
\centering
\begin{tabular}{cc}
    \includegraphics[width=0.45\textwidth]{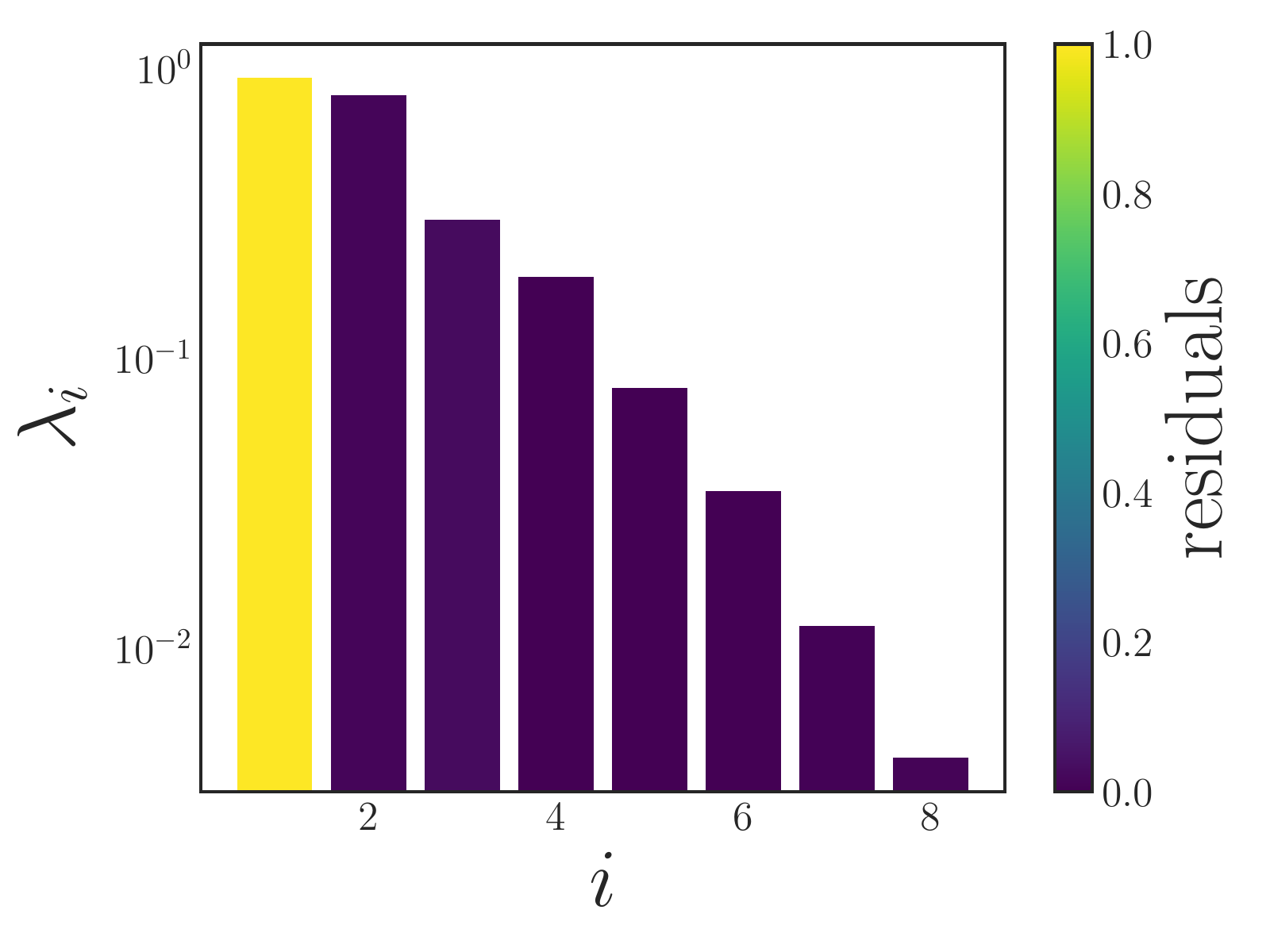}  &
    \includegraphics[width=0.45\textwidth]{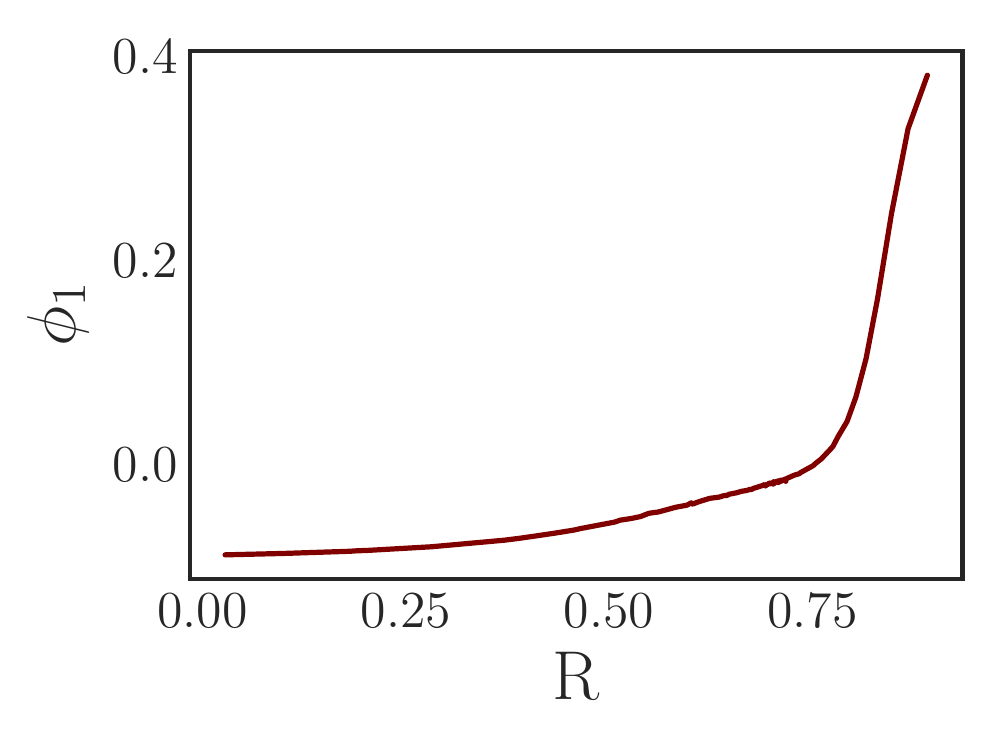} \\
    (a) & (b)
\end{tabular}
    \caption{(a) Eigenvalues of the diffusion map operator on the oscillator density dynamic data, sorted in decreasing order and colored by the corresponding residual value produced by the local linear regression algorithm. The first eigenfunction $\phi_{1}$ suffices to parametrize the behavior.
    (b) A plot demonstrating the correspondence between the discovered diffusion map coordinate $\phi_{1}$ and the magnitude of the typical Kuramoto order parameter $R$. 
    It clearly shows the one-to-one correspondence between the two, verifying that the diffusion map coordinate is an ``equivalent" coarse variable to $R$ for this system.}
    \label{fig:R_dmap}
\end{figure}

Using time series of density data instead of individual phase data simplifies the comparison of different snapshots, and thus facilitates the selection of the kernel in the DMAPs algorithm. For this we select the standard Gaussian kernel with the Euclidean metric for the comparison of our oscillator density vectors ($d \in \mathbb{R}^{200}$) with diffusion map tuning parameters of $\alpha=1$ selecting the Laplace-Beltrami operator, and $\epsilon\approx 1.54$ as the kernel bandwidth parameter. This results in a single diffusion map coordinate (eigenfunction) that is deemed significant by the local linear regression method, see \textbf{Fig.~\ref{fig:R_dmap} (a)}.

Comparing this diffusion map coordinate $\phi_{1}$ to the Kuramoto order parameter $R$ reveals there is a one-to-one correspondence between the two, as \textbf{Fig.~\ref{fig:R_dmap} (b)} clearly illustrates: the plot is monotonic, and its derivative remains bounded away from zero and from infinity. This means that the diffusion map coordinate contains the same information as the established Kuramoto order parameter, and therefore constitutes a suitable coarse variable for this system. This is particularly interesting as we managed to identify this variable directly through manifold learning on the minimally pre-processed phase data, without referencing Kuramoto's order parameter. In this way, we have demonstrated a process that ``learns" bespoke coarse variables, circumventing the traditional discovery/invention process.

While this process is both systematic and automated, we must point out that the key step in this process, the choice of the relevant features of the data, still requires a modicum of insight. Different choices of the relevant features of the data, either through pre-processing and/or selection of the kernel, will produce different coarse variables, that may well be ``reconcilable," i.e. one-to-one with each other and with $R$. 
\section*{Learning the Dynamic Behavior of Data-Driven Coarse Variables with Neural Networks}
\label{sec:ode_learning}
Once descriptive coarse variables have been identified, it is desirable to find descriptions of their behavior (evolution laws, typically in the form of ordinary or partial differential equations, ODEs or PDEs). However, finding analytical expressions for these descriptions (``laws") is often extremely challenging, if not practically infeasible, and may rely on \textit{ad hoc} methods. The process used to learn the analytical equations that describe the behavior of the Kuramoto order parameter in the continuum (infinite oscillator) limit exemplifies these types of difficulties \citep{ott2008low}. As a result of this process, Ott and Antonsen showed that the continuum limit of the all-to-all coupled Kuramoto model \eqref{eq:simple_kuramoto} admits an attracting, invariant manifold. For Cauchy distributed frequencies, the conservation PDE governing the oscillator density can be analytically transformed into a pair of ODEs that describe the time evolution of the Kuramoto order parameter along this manifold,
\begin{equation}
\begin{aligned}
    \frac{dR}{dt}&=-\gamma R + \frac{KR}{2} (1-R^{2}) \label{eq:Ott_Antosen_ODEs} \\
    \frac{d\psi}{dt}&=0,
\end{aligned}
\end{equation}
where $R$, $\psi$, $K$, and $\gamma$ are the phase coherence, average phase, coupling constant, and Cauchy distribution parameter respectively with $t$ as time. As this manifold is invariant and attracting, the dynamics of the order parameter can be accurately described by these equations after a short initial transient.

In the following sections, we present an alternative, data-driven approach to learning the behavior of coarse variables directly from time series of observational data. As part of this approach we make use of a recurrent neural network architecture ``templated" on numerical time integration schemes, which allows us to learn the time derivatives of state variables from flow data in a general and systematic way. We illustrate this approach through an example in which we learn the aforementioned ODEs \eqref{eq:Ott_Antosen_ODEs} that govern the behavior of the Kuramoto order parameter in the continuum limit from data. We then compare the result of our neural network based approach to standard finite differences complemented with geometric harmonics. Throughout this example we only observe the magnitude of the order parameter $R$ and neglect the angle $\psi$, as the magnitude captures the relevant synchronization dynamics of this model.
\subsection*{A Neural Network Based on a Numerical Integration Scheme}
Numerical integration algorithms for ODEs rely on knowledge of the time derivative in order to approximate a future state. If an analytical formula for the time derivative is not available or unknown, it can be approximated from time series of observations through, say, the use of finite differences, with known associated accuracy problems, especially when the data is scarce. Here, we discuss an alternative, neural network based approach to learning time derivatives (the ``right hand sides" of ODEs) from discrete time observations.

Artificial neural networks have gained prominence for their expressiveness and generality, and especially for their ability to model nonlinear behavior. These qualities have led to their widespread adoption and use in areas as diverse as image \citep{simonyan2014very} and speech recognition \citep{graves2013speech}, financial modeling and prediction \citep{guresen2011using}, and general game playing \citep{silver2016mastering, vinyals2017starcraft}. For the approximation of dynamical systems, neural networks have been used for tasks such as the accurate approximation of functions and their derivatives \citep{cardaliaguet1992approximation}, system identification and control \citep{narendra1990identification, subudhi2011differential}, and system modeling \citep{chow2000modeling}. A variety of network architectures have been studied for the analysis of dynamical systems, such as feedfoward networks \citep{rico1992discrete}, recurrent networks \citep{chow2000modeling}, high-order networks \citep{kosmatopoulos1995high}, and multistep networks \citep{raissi2018multistep}, along with novel training approaches such as the differential evolution approach \citep{chow2000modeling}. Here we focus on a feedforward stepping approach, whose utility for accurately modeling system dynamics has been previously demonstrated \citep{rico1992discrete, raissi2018multistep}.

This feedfoward approach is a method for approximating the functional form of the right-hand-side of systems modeled through autonomous ODEs \citep{rico1992discrete, rico1993continuous, rico1995nonlinearODE}. The crux of the approach is to approximate the time derivative of the system with a feedforward (and here, a recurrent) neural network. This neural network approximation can then be used in place of a first-principles based right-hand-side in any initial value solver, such as the Euler or Runge-Kutta methods, to produce a new, neural network based time-stepper \citep{rico1995nonlinearODE}. In other words, a neural network architecture templated on a numerical time-stepping processes can be trained to learn an approximation of the right-hand-side of the system equations from pairs of inputs and outputs, where the input is the state $y(t)$ at time $t$ and the output is $y(t+\Delta t)$ at time $t+\Delta t$. By training such a neural network architecture on pairs of state variable observations, $(y(t), y(t+\Delta t))$, a part of the neural network learns an approximation of the right-hand-side of the evolution equation. This is a surrogate model, which subsequently, using any good integration algorithm, can produce a flow that closely matches the training data. Following successful training, an estimate of the right-hand-side of the evolution equation for any out-of-sample initial system states can be easily accessed by evaluating the neural network directly for these system states.

In addition to its application to learning the right-hand-sides of systems of ODEs, this type of neural network architecture can also be extended to learn the right hand sides of PDEs discretized as systems of ODEs through a method of lines approach \citep{gonzalez1998identificationPDE}. While any explicit integration algorithm that only requires knowledge of the right-hand-side can be used to devise acceptable neural network architectures for learning unknown evolution equations, we will focus here on the fourth order Runge-Kutta algorithm for our analysis and computations.

\begin{figure}[h]
    \centering
    \includegraphics[width=0.8\textwidth]{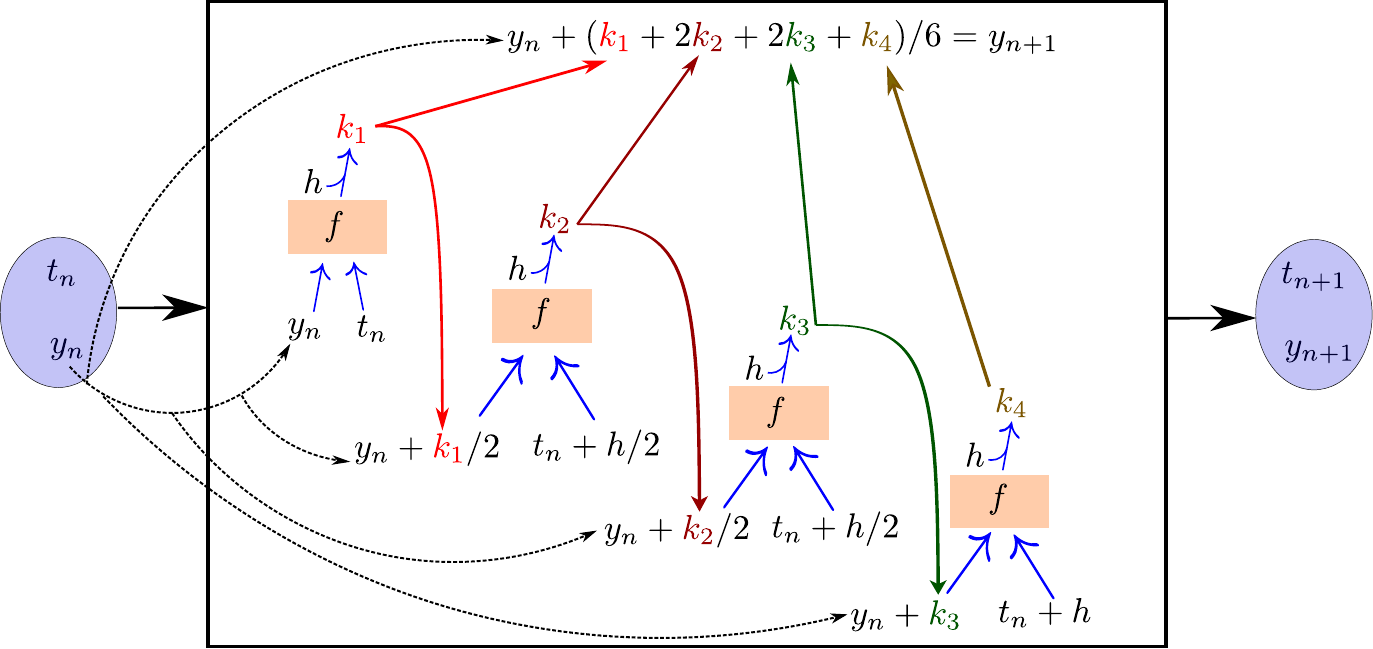}
    \caption{The neural network approach for learning unknown ODE right-hand-sides templated on a fourth order, fixed-step Runge-Kutta numerical integration algorithm. The neural ``sub"-network $f$ is reused for each stage of the integration algorithm in order to compute the loss function and train the network.}
    \label{fig:RK4_scheme}
\end{figure}

A schematic of the feedforward recurrent neural network architecture we construct templated on a fourth order, fixed-step Runge-Kutta algorithm is illustrated in \textbf{Fig.~\ref{fig:RK4_scheme}}. An important feature of this method with the Runge-Kutta algorithm is that the ``black box" neural network evaluating $f$ (a recurrent component of the overall network architecture) is shared between all of the stages of the algorithm ($k_{1}, k_{2}, k_{3}, k_{4}$). That is, there is a single copy of the neural ``sub"-network $f$ that is evaluated multiple times per time step with different inputs as required by the Runge-Kutta algorithm in order to produce the output.

In the following section, we use this Runge-Kutta scheme to learn the ODE governing the behavior of the magnitude of the Kuramoto order parameter $R$ in a data-driven way.
\subsection*{Generating the Flow (Training) Data}
As discussed in the previous section, we need to generate pairs of flow data of the coarse variable, $(R(t), R(t+\Delta t))$, in order to train the neural network based approach. We begin by considering 2000 unique (different initial phases and frequencies) simulations of the simple all-to-all coupled Kuramoto model \eqref{eq:simple_kuramoto} with 8000 oscillators ($N=8000$) with a coupling constant $K=2$, and independently sampled frequencies $\omega_{i}$ drawn from a Cauchy distribution with $\gamma=0.5$.

\begin{figure}[h]
\centering
\begin{tabular}{cc}
\includegraphics[width=0.45\textwidth]{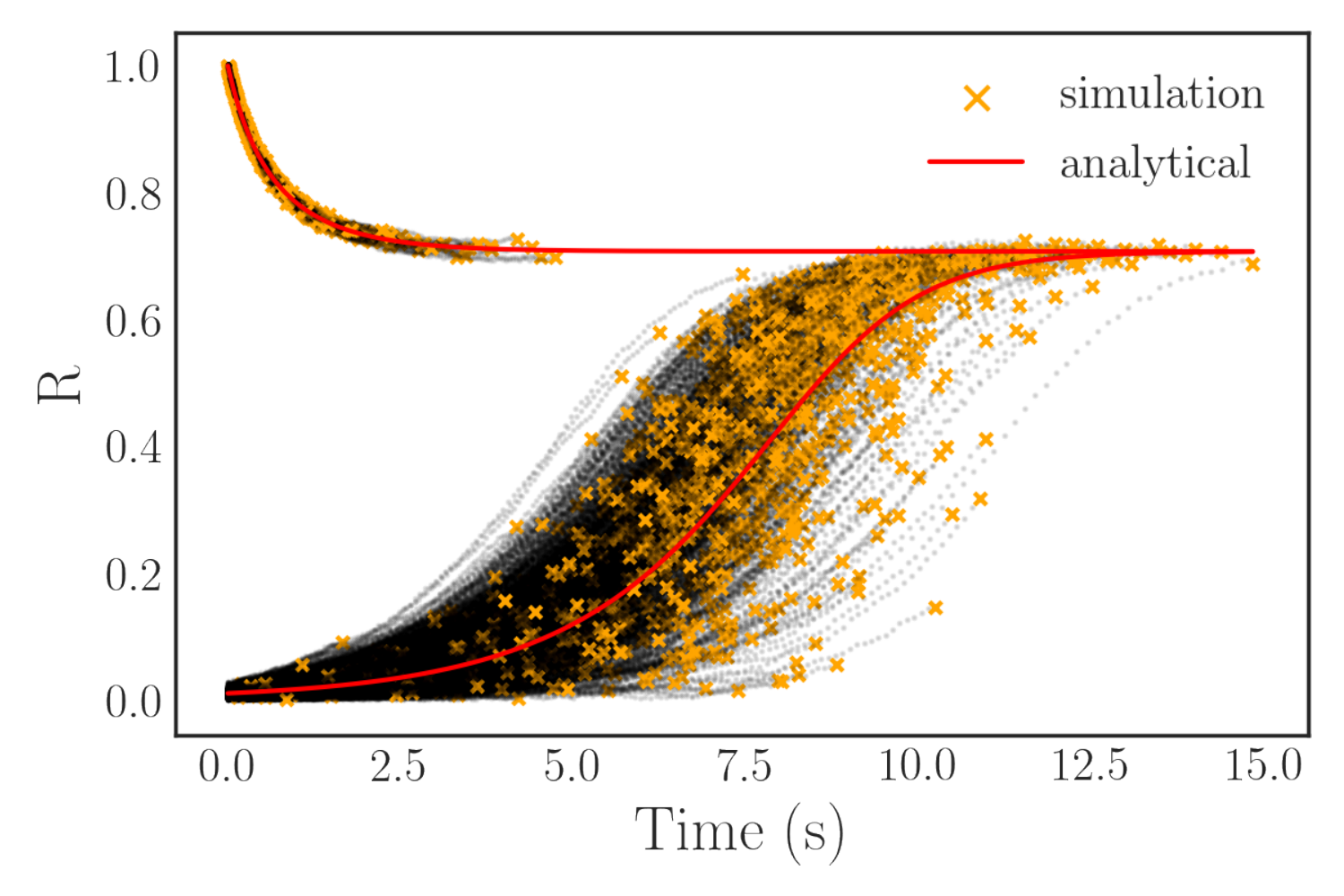} & 
\includegraphics[width=0.45\textwidth]{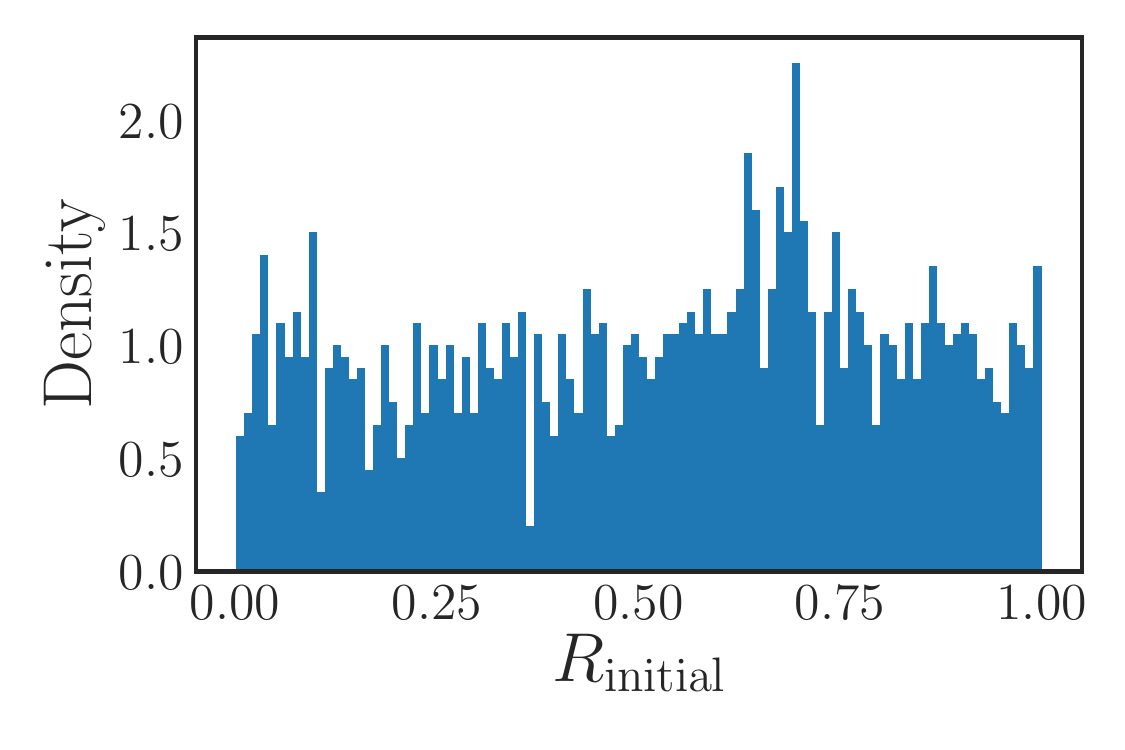} \\
(a) & (b) 
\end{tabular}
    \caption{(a) A plot of our initialization trajectories (in black) versus the expected continuum limit solution (in red). The end point of each black trajectory, denoted by an orange star, corresponds to a value of $R(t)$ that is used as a starting point for the flow training data. The noise evident in the simulated trajectories is caused by both randomness in the frequencies (each trajectory is a different sample from the distribution), and randomness in the arrangement of the oscillators with respect to their frequencies (only for the $R\approx 0$ starting condition). (b) The approximately uniform sampling of the initial $R$-values $R(t)$ used for flow generation. Note that a uniform distribution is not required, however it is desirable to facilitate training. It is only necessary that the values cover the range of $R$-space in order to allow us to learn the ODE over the entire $[0, 1]$ domain.}
    \label{fig:Initialization Trajectories}
\end{figure}

One of the difficulties of generating flow data for the order parameter is adequately sampling the entire range of $R \in [0, 1]$. As the order parameter is not a quantity that we are directly simulating, and instead depends on the phases of the oscillators in a complex way, it is difficult to consistently initialize flows to arbitrary values of $R$. We surmount this difficulty by means of an initialization integration approach. This initialization approach consists of initializing each of the unique oscillator simulations to easily expressed order parameter values, either $R\approx 0$ (uniformly distributed random initial phases on $[0, 2\pi]$) or $R=1$ (identical initial phases), and then integrating them for different amounts of time to produce a set of 2000 variegated starting points for our $R$-flow data, $R(t)$. These points are the final points of the black initialization trajectories shown in \textbf{Fig.~\ref{fig:Initialization Trajectories} (a)} and are marked by orange stars. By judiciously selecting the integration times, it is possible to produce a set of initial points $R(t)$ that covers the entire range of possible $R$-values nearly uniformly, see \textbf{Fig.~\ref{fig:Initialization Trajectories} (b)}.

After generating the starting points of the flows $R(t)$ with this initialization approach, we proceed to produce the corresponding end points $R(t+\Delta t)$ through a detailed time integration of the oscillator phases. This process consists of first integrating the oscillator phases corresponding to $R(t)$ forward in time for a chosen time step $\Delta t$, and then computing the order parameter for the resulting new phases. We select two different reporting time horizons for this integration, $\Delta t=0.01$, and $\Delta t=0.05$, to produce two different sets of flow data to assess the sensitivity of our method to the flow time. For all of these simulations, we use Scipy's Runge-Kutta integrator (\textit{solve\_ivp} with the RK45 option) to perform the time integration.

This entire process results in two collections of training data, each consisting of 2000 pairs of the form $(R(t), R(t+\Delta t))$, with one collection for each choice of $\Delta t$. Each of these collections is used individually in the following section to train our neural network to approximate the right-hand-side of the coarse evolution equation for $R(t)$.
\subsection*{Training the Neural Network Scheme}
Our neural network architecture is based on a standard fixed step-size, fourth order Runge-Kutta integration algorithm complemented with a feedforward neural network with 3 hidden layers of 24 neurons each, \textbf{Fig.~\ref{fig:RK4_scheme}}. We initialize our network with a uniform Glorot procedure \citep{glorot2010understanding} for the kernels, and zeros for the biases. For training, we use TensorFlow's Adam optimizer \citep{kingma2014adam} with a learning rate of $10^{-3}$ and a mean squared error loss on the final flow point $R(t+\Delta t)$ with full batch training (all of the training data used in each training epoch).

\begin{figure}[!htp]
\centering
\begin{tabular}{cc}
\includegraphics[width=0.45\textwidth]{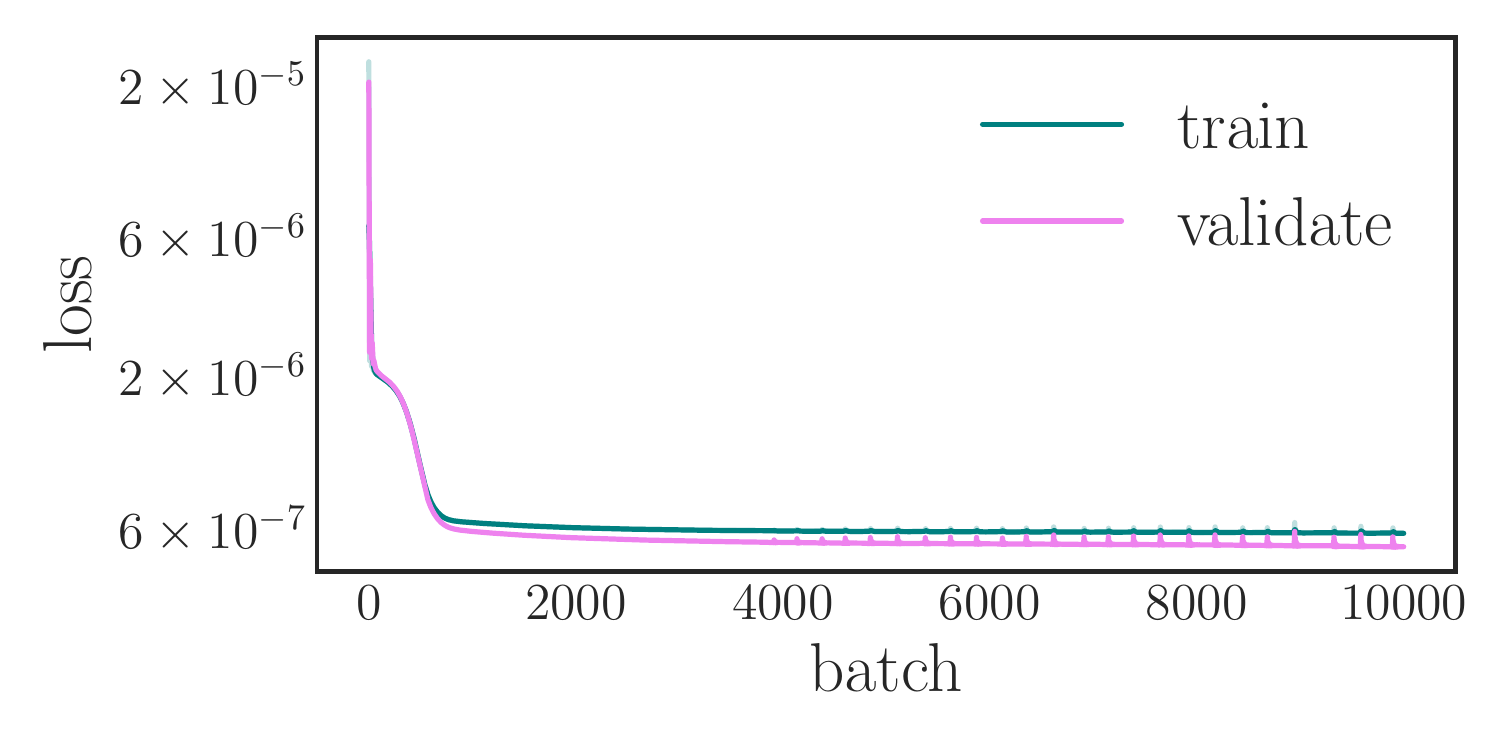} &
\includegraphics[width=0.45\textwidth]{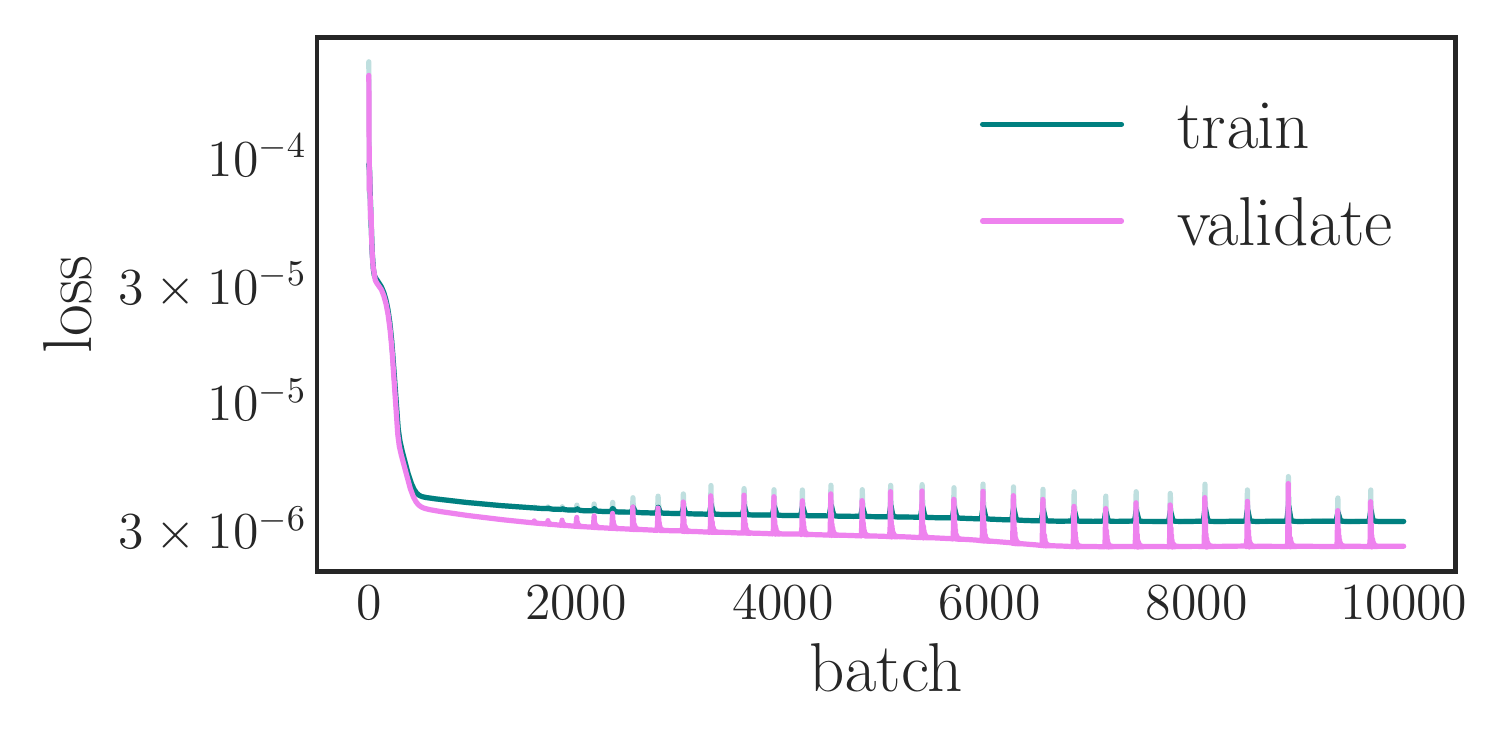} \\
(a) & (b) \\
\multicolumn{2}{c}{\includegraphics[width=0.70\textwidth]{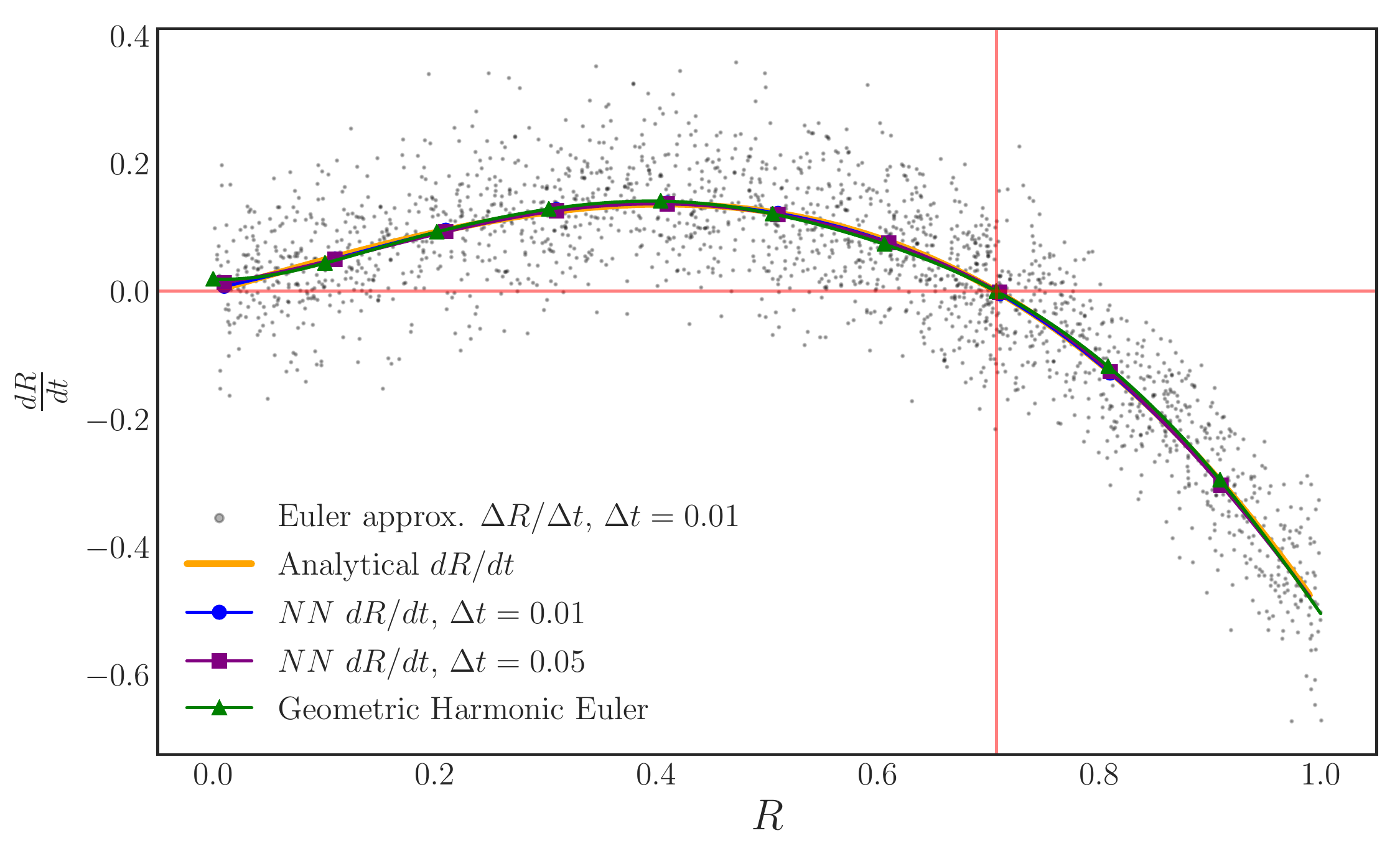}} \\
\multicolumn{2}{c}{(c)} \\
\multicolumn{2}{c}{\includegraphics[width=0.70\textwidth]{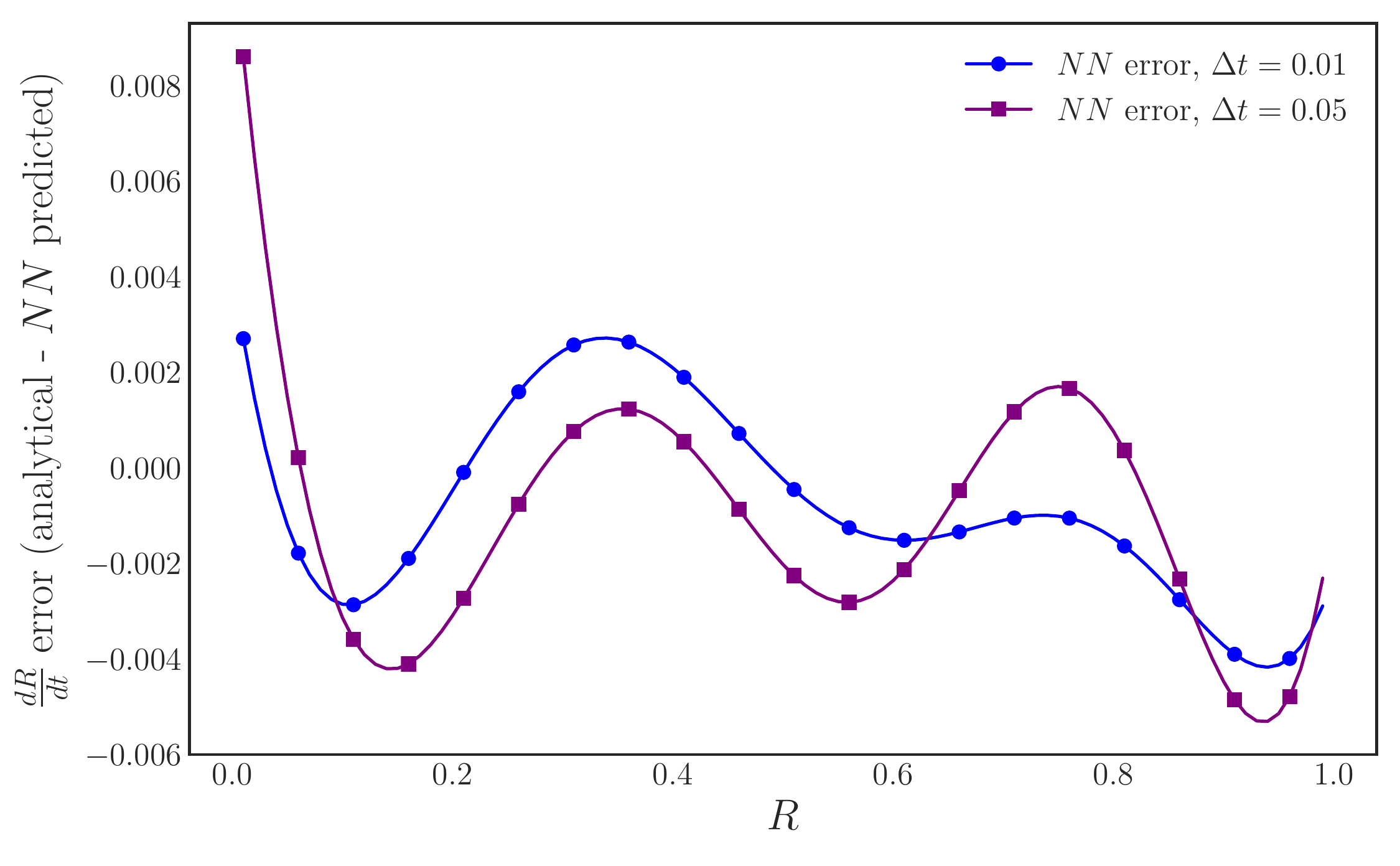}} \\
\multicolumn{2}{c}{(d)}
\end{tabular}
    \caption{The training and validation loss of the neural network training for: (a) the $\Delta t=0.01$ time horizon, and (b) the $\Delta t=0.05$ time horizon. (c) Plots of the learned, neural network right-hand-side for two different time horizons (blue, purple) and the analytically derived right-hand-side (orange). A forward Euler approximation (black) and its geometric harmonic interpolation (green) are included for comparison. (d) The error behaves similarly for both time horizons, with a magnitude less than 0.01 for the entirety of the domain.}
    \label{fig:NN_results}
\end{figure}

We train one neural network for each time horizon data set, $\Delta t =0.01, 0.05$, to investigate the sensitivity of the right-hand-side estimation to the flow time. In each case we use a split of 10\% of the data for validation and 90\% of the data for training to check for overfitting and proceed to train the network with full batch training (batches of 1800 data points) until we reach a sufficiently small loss value (10,000 epochs in total), \textbf{Fig.~\ref{fig:NN_results} (a, b)}. As \textbf{Fig.~\ref{fig:NN_results} (c, d)} shows, the different time steps behave similarly with minimal deviation from the analytical values, indicating a low sensitivity to the time step for this range of values. The gray points in \textbf{Fig.~\ref{fig:NN_results} (c)} are forward Euler approximations of the time derivative of the order parameter that are computed from the flow data generated with $\Delta t=0.01$. These points reveal the scatter in the data, as noted in \textbf{Fig.~\ref{fig:Initialization Trajectories} (a)}, and illustrate the power of the neural network time-stepping process to regress the data and average out the noise to find what can be thought of as an expected value (over oscillator ensemble realizations) of the time derivative. To provide a comparison between finite difference time-derivative estimates and our neural network based approach, we compute a geometric harmonic interpolation of the Euler approximations with 10 geometric harmonics and include the result in \textbf{Fig.~\ref{fig:NN_results} (c)}.

\begin{figure}[h]
    \centering
    \includegraphics[width=0.8\textwidth]{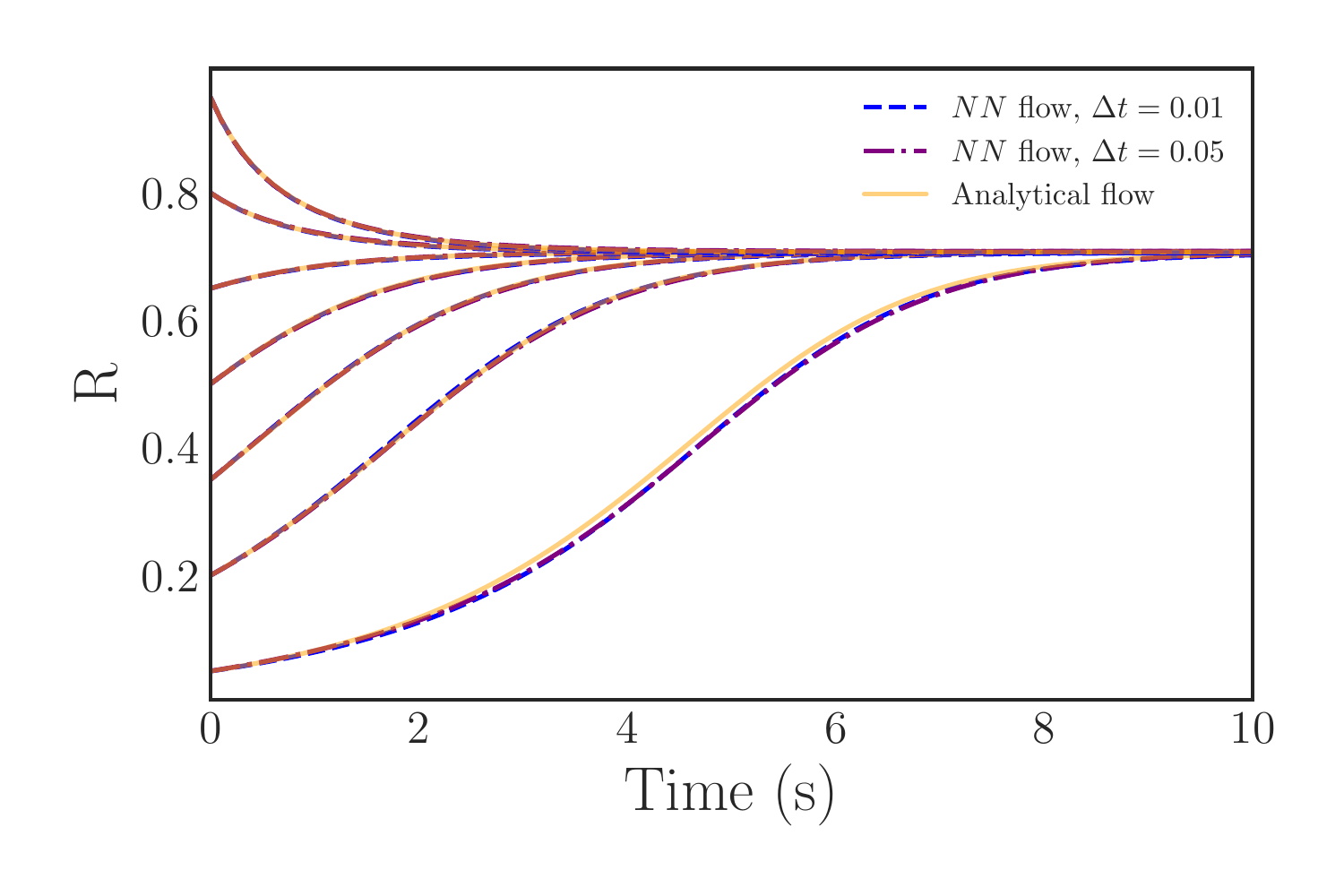}
    \caption{Plots of trajectories generated from the learned, neural network, evolution law right-hand-side (blue, purple) versus the analytically known evolution law (orange). The learned trajectories exhibit nearly identical behavior to the analytical trajectories and, importantly, converge to the correct steady state.}
    \label{fig:NN trajectories}
\end{figure}

As an added point of comparison between our neural network right-hand-sides and the analytical expression, we integrate flows for a variety of initial conditions with both of the neural network approximated right-hand-sides and the analytical (theoretical, infinite oscillator limit) right-hand-side. As \textbf{Fig.~\ref{fig:NN trajectories}} illustrates, the neural network equations produce flows that both closely match the analytical solution over the majority of the $R$-domain and converge to the correct steady state. Thus, the neural network approach successfully learns an accurate approximation of both the behavior and the governing ODEs of the Kuramoto order parameter in the continuum limit.

One of the greatest utilities of this approach is its generality, which enables it to be applied to cases in which analytical approaches have not yet been devised, or may not be practical. Throughout the previous example we assumed knowledge of the coarse variable, the Kuramoto order parameter, and learned its governing ODEs. However, it is important to realize that this same technique can also be applied to our ``discovered'' coarse variables, that we identified earlier in this paper through manifold learning techniques. In this way, this methodology allows one to \textbf{both} identify the descriptive coarse variables \textbf{and} learn their behaviors (ODEs) in a general, data-driven way that is tailored to the specific problem by the nature of the technique. We conclude this section with a demonstration of such an application.

\begin{figure}[h]
    \centering
    \includegraphics[width=0.6\textwidth]{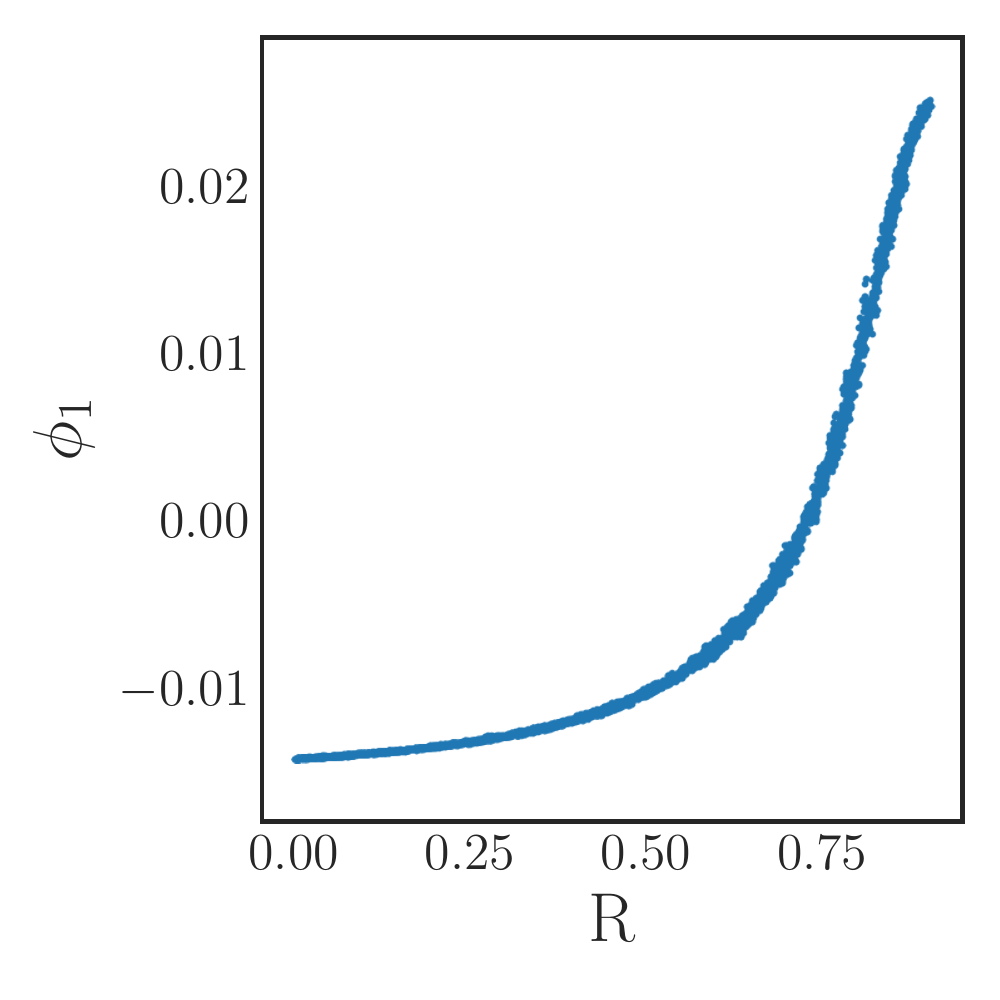}
    \caption{The coarse variable identified by our manifold learning procedure applied to data generated in this section. Only the single coarse coordinate $\phi_{1}$ was deemed significant by the local linear regression method. Even in the presence of noisy data, there is nearly a one-to-one map between the discovered coordinate $\phi_{1}$ and the analytical order parameter $R$.}
    \label{fig:Learned_R_Full_Data}
\end{figure}

Following the same methodology outlined in the ``Order Parameter Identification'' section, we combine the phase data corresponding to the two $(R(t), R(t+\Delta t))$ data sets used earlier in this section and then bin the oscillator phases for each time point to produce oscillator density data points. As before, we use this oscillator density data as the input to the DMAPs algorithm with a Gaussian kernel with the Euclidean distance and find that it is described by a single significant eigenfunction, as determined by the local linear regression method. \textbf{Fig.~\ref{fig:Learned_R_Full_Data}} illustrates the nearly one-to-one mapping between this discovered diffusion map coordinate and the analytical order parameter $R$. Following our discovery of a coarse variable, we then use the neural network process outlined earlier in this section to approximate the evolution law (the right-hand-side of the unknown ODE) for our discovered, diffusion map coarse variable $\phi_{1}$.

\begin{figure}[h]
    \centering
    \includegraphics[width=1.0\textwidth]{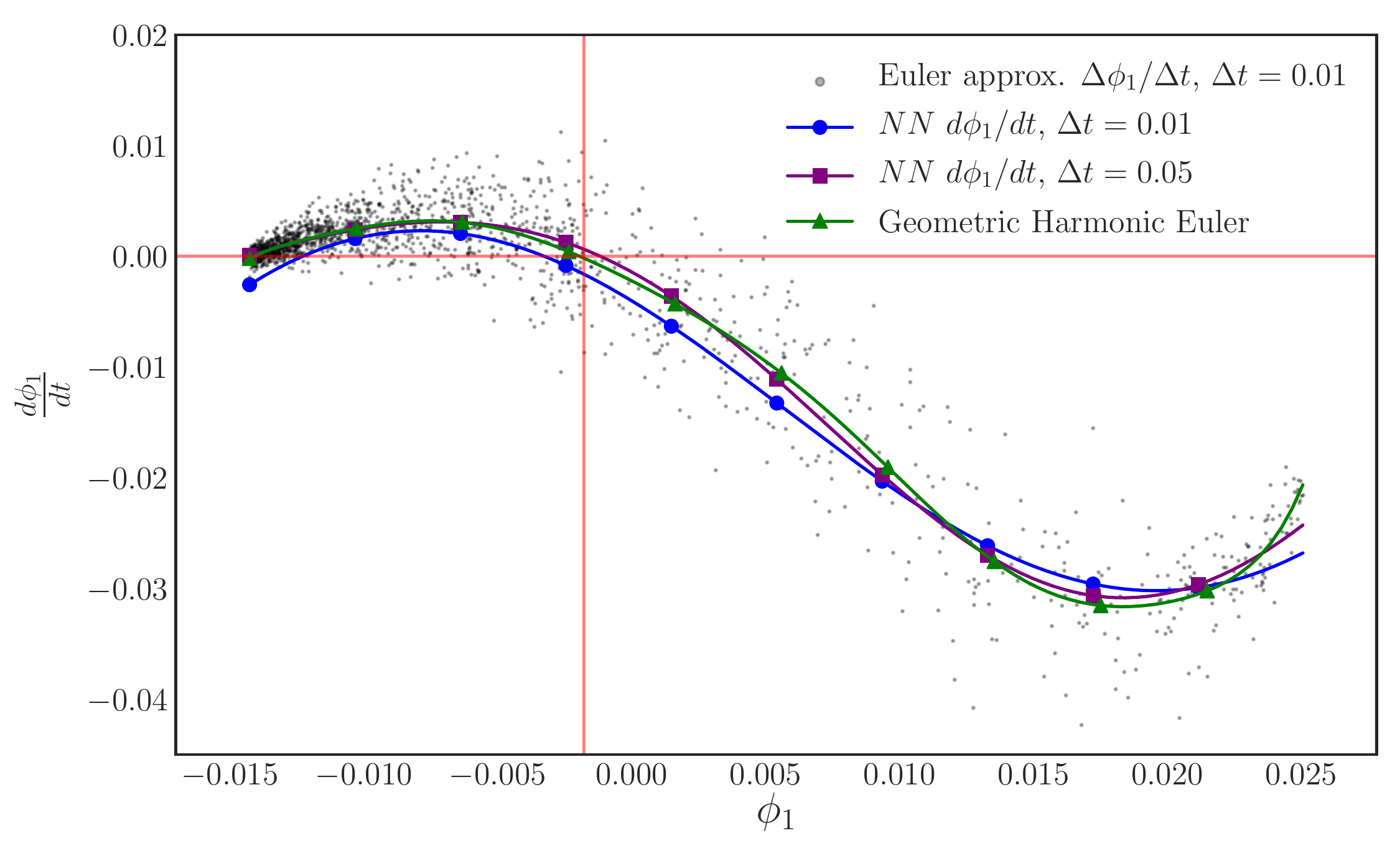}
    \caption{The learned evolution law (ODE right-hand-side) of our discovered order parameter $\phi_{1}$ for each reporting time horizon. The steady state of this system in these coordinates is highlighted in red. A forward Euler approximation (black) and its geometric harmonic interpolation is included to provide a point of comparison and to highlight the scatter in the data.}
    \label{fig:Learned_R_time_derivative}
\end{figure}

By keeping track of the the $\phi_{1}$ values corresponding to the $R$ values for each data set, we form two sets of pairs of training data as before, $(\phi_{1}(t), \phi_{1}(t+\Delta t))$, with one for each value of $\Delta t$, but now in the manifold learning derived variable $\phi_{1}$. Using an identical procedure to the one used to learn the right-hand-side of the $R$ equation, we define a feedforward neural network with 3 hidden layers of 24 neurons each and use it to approximate the time derivative of $\phi_{1}$ in an architecture templated on a fixed step-size, fourth order Runge-Kutta algorithm. As before, we initialize this neural network integration procedure with a uniform Glorot procedure for the kernels, and zeros for the biases. We define the loss to be the mean squared error on the final point of the training flow and then train the network with full batch training with the Adam optimizer with a learning rate of $10^{-3}$. \textbf{Fig.~\ref{fig:Learned_R_time_derivative}} illustrates the result of this training procedure for each time step. As before, we include the forward Euler approximation of the $\phi_{1}$ time derivative for $\Delta t=0.01$ and its geometric harmonic interpolation with 10 geometric harmonics as a point of reference.

Our recurrent, integrator-based neural network architecture, appears to successfully learn the evolution equation for $\phi_{1}$ over the domain. However, the fit does not appear to be as accurate as the one found for $R$ itself. Furthermore, this training required 50,000 epochs to reach this level of accuracy compared to the 10,000 used for $R$. We argue that both the difficulty in training and the lower quality of this fit can be attributed to the bias in sampling introduced by the nonuniform sampling in our diffusion map coordinate, demonstrated by the accumulation of the black data points in \textbf{Fig.~\ref{fig:Learned_R_time_derivative}}. Earlier we noted that we took great care to produce training data sets that sample the $R$ domain in a nearly uniform fashion, \textbf{Fig.~\ref{fig:Initialization Trajectories} (b)}. As demonstrated by \textbf{Fig.~\ref{fig:Learned_R_Full_Data}}, the mapping from $R$ to $\phi_{1}$ is one-to-one, but does not preserve the shape of the sampling density due to its nonlinearity, leading to a bias in the sampling of the $\phi_{1}$ domain. However, despite this bias, the neural network procedure manages to learn a time derivative near the visual average of its observed Euler approximation.

To summarize, we have shown how our order parameter identification method can discover a coarse variable that shows good agreement (i.e. is one-to-one) with a known analytical variable, even in the case of non-ideal, noisy data. Furthermore, we have demonstrated that even with biased sampling, our recurrent, integrator templated neural network architecture is capable of successfully learning the right-hand-side of the evolution of our discovered coarse variable, effectively smoothing its Euler finite difference estimates, and certainly closely approximating the correct steady state.
\section*{Identifying ``Effective'' Parameters}
\label{sec:effective parameters}
Up to this point we focused on coarse-graining (reducing the dimension) the system state variables required to formulate an effective dynamic coupled oscillator model. Often, however, complex systems whose dynamics depend explicitly on several parameters can also admit \textit{a reduced set of parameter combinations} (or ``effective'' parameters) that depend on the full set of parameters in complex, nonlinear ways. Finding these reduced sets of parameters often requires insight, along with trial and error. Here, we present a methodology for discovering such \textit{effective parameters} in a data-driven way via a modification of diffusion maps, our manifold learning technique of choice in this paper \citep{holiday2019manifold}. A strong motivation for this work comes from the determination of explicit dimensionless parameters from the (possibly long) list of dimensional parameters of a physical model.

Throughout this section we investigate variations of the Kuramoto model and show how DMAPs can be used to discover effective parameters that accurately describe the phases of Kuramoto oscillators when synchronized (at steady state in a rotating frame). The general idea is to use an \textit{output-only informed kernel} for the diffusion map. With this approach, we only consider the outputs of the system, here steady state phase data, as the input to the DMAPs algorithm/kernel computation and neglect the values of the oscillator parameters, which embody the heterogeneity of the oscillator ensemble. This allows us to discover the intrinsic dimensionality of the state space at synchronization, independent of the detailed list of system parameters. The significant eigenfunctions provided by DMAPs, as determined by the local linear regression method, serve as new coordinates for the state space, that is, their values completely determine the values of the state variables at steady state \eqref{eq:dmap_analysis}.
\begin{alignat}{2}
\label{eq:original_analysis}
&\theta_{\infty}(\textbf{p})
\quad &&\text{Original analysis}\\
\label{eq:dmap_analysis}
&\theta_{\infty}(\boldsymbol{\phi})
\quad &&\text{Diffusion map coordinate(s)}\\
\label{eq:effective_parameters}
&\mathbf{p} \mapsto \boldsymbol{\phi}
\quad &&\text{New ``effective'' parameter(s)}
\end{alignat}
Since these effective coordinates describe the variability of the steady state variables (which depend on the detailed system parameters), the eigenfunctions themselves provide a new, data-driven set of effective parameters for the model. If fewer eigenfunctions are required to describe the steady state space than there are original parameters, then these eigenfunctions furnish an \textit{effective reduced set of parameters}. Once such data-driven effective parameters are found, they can be compared to the original parameters through regression to determine how they are related \eqref{eq:effective_parameters}. We illustrate this methodology through a series of examples.
\subsection*{A Simple Example: the Kuramoto Model with Heterogeneous Coupling Coefficients}
We begin by considering a simple variation of the Kuramoto model in which we include the coupling constant $K_{i}$ as an oscillator heterogeneity in addition to the frequencies $\omega_{i}$. The governing equations for this model are
\begin{equation}
\frac{d\theta_i}{dt}=\omega_i+\frac{K_{i}}{N}\sum_{j=1}^N \sin(\theta_j-\theta_i), \quad \text{for}\ i=1,\dots, N.
\label{eq:kuramoto_hetero_coeffs}
\end{equation}
Similar to the typical Kuramoto model, the phase synchronization of the oscillators can be described with the usual complex-valued order parameter
\begin{equation}
Re^{i\psi}=\frac{1}{N}\sum_{j=1}^N e^{i\theta_j}.
\end{equation}
As with the the typical Kuramoto model, this variation admits a steady state in a rotating frame if there is complete synchronization.

Our goal for this model is to show that even though there are two model heterogeneities, ($\omega_{i}, K_{i}$), the steady state phases can be described by a single ``effective'' parameter. That is, we will demonstrate that there is a new parameter that depends on both of the original two parameters, such that the steady state oscillator phases only depend on this single combination of the original parameters.

In order to find this effective parameter, we apply the DMAPs algorithm to the steady state phase data with an \textit{output-only informed kernel}. This means that our observations consist solely of the output of the model, here the steady state phases of the oscillators $\theta_{i, \infty}=\theta_{i}(t_{\infty})$ (where $t_{\infty}$ is a time after which the oscillators have reached a steady state), and ignore the oscillator heterogeneities $(\omega_{i}$, $K_{i})$. We use the eigenfunctions provided by the DMAPs algorithm to define a change of variables for the parameter space $(\omega,K)$. As we show, only a single eigenfunction $\phi_{1}$ is required to describe the phase data. Thus, this change of variables is a many-to-one map and provides a reduction from the two original parameters $(\omega, K)$ to the single effective parameter, $\phi_{1}$.
\begin{align}
(\omega,K) & \mapsto \theta_{\infty} \in \mathbb{R}
\quad \text{Original analysis} \\
\phi_{1}(\omega,K) & \mapsto \theta_{\infty} \in \mathbb{R}
\quad \text{Diffusion map coordinate}
\end{align}

For our simulations we consider 1500 oscillators ($N=1500$) with uniformly randomly distributed initial phases over $[0, 2\pi]$, uniformly randomly distributed frequencies over $[-\pi, \pi]$, and uniformly randomly distributed coupling coefficients over $[10, 100]$. We select these parameter values as they lead to complete synchronization, and hence a steady state in a rotating frame. We integrate the oscillators with Scipy's vode integrator until they achieve complete synchronization, and then transform the phases into a rotating frame.

Next, we apply the DMAPs algorithm to the steady state phases with a Gaussian kernel with the Euclidean distance and parameters of $\alpha=1$ for the Laplace-Beltrami operator and $\epsilon=0.5$ for the kernel bandwidth parameter. The kernel is given by
\begin{equation}
K_{ij} = \exp\left( - \frac {\Vert x_i-x_j \Vert_2^2}{\epsilon^2}\right)\quad i,j=1,\dots,1500,
\end{equation}
where $\Vert \cdot \Vert_2$ is the Euclidean norm in the complex plane\footnote{It is necessary to map the oscillator phases to the complex plane to avoid the unfortunate occurrence of the phases lying across the branch cut of the multi-valued argument function. Taking the Euclidean norm of the difference between oscillator phases in this situation would result in the DMAPs algorithm incorrectly identifying two different clusters of oscillators split across the branch cut instead of a single group. By first mapping the phases to the complex plane and then computing the distances there, we avoid this possibility and correctly identify a single group of oscillators.} of 1500 oscillator phases, e.g.~$x_j = e^{i\theta_j(t_{\infty})}$.

\begin{figure}[h]
    \centering
    \includegraphics[width=1.0\textwidth]{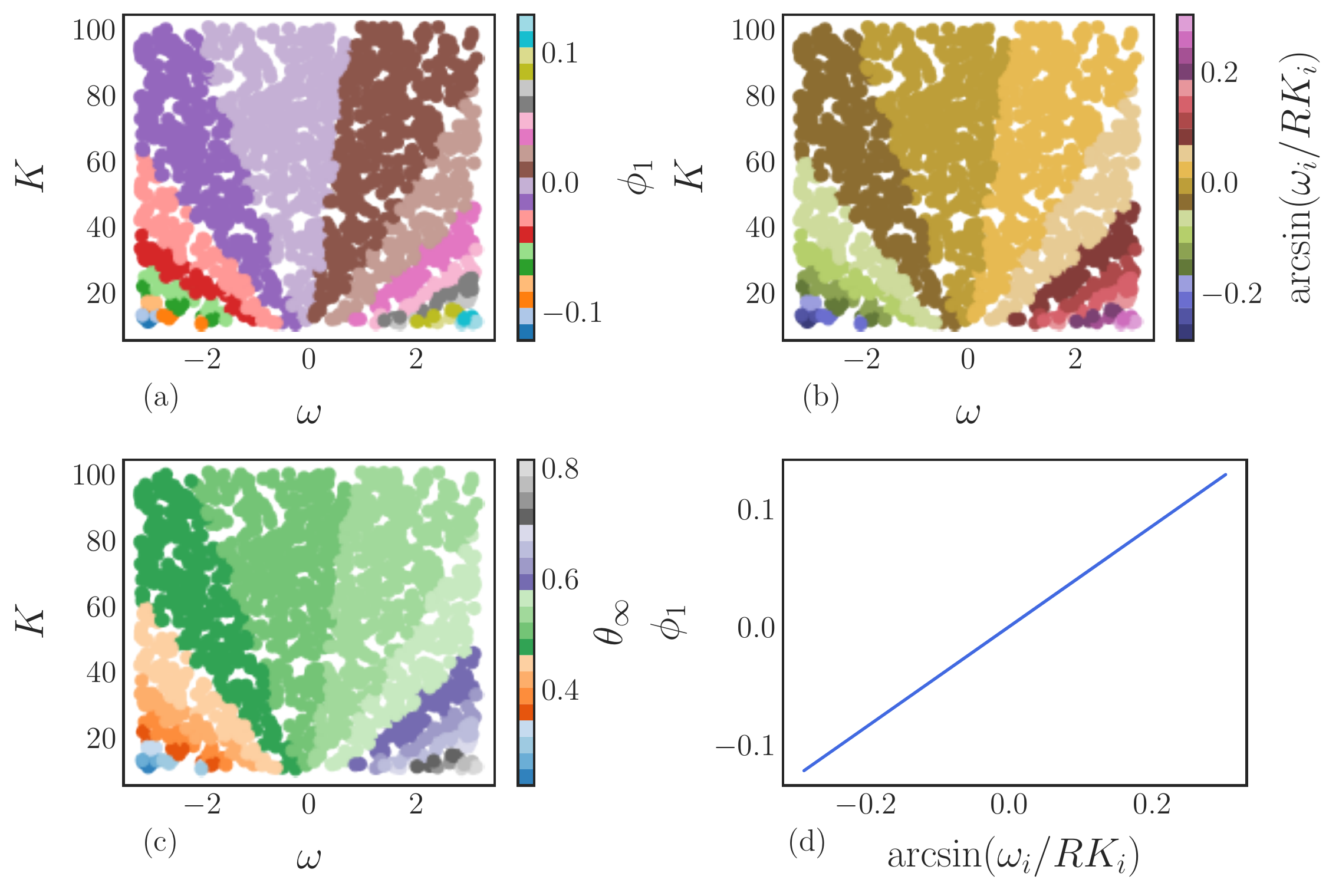}
    \caption{
    (a) The parameter space $(\omega, K)$ colored by the diffusion map coordinate $\phi_{1}$.
    (b) The parameter space $(\omega, K)$ colored by the analytical parameter \eqref{eq:closed_form_eff}.
    (c) The parameter space $(\omega, K)$ colored by the steady state phase $\theta_{\infty}$.
    (d) A comparison between the data-driven parameter and the analytically derived effective parameter. The one-to-one mapping between the two verifies their equivalence.}
    \label{fig:simple_kuramoto_parameter}
\end{figure}

We use the local linear regression method to verify that only a single diffusion map eigenfunction $\phi_{1}$ is required to represent this phase data, resulting in a single, data-driven effective parameter. A coloring of the original, two-dimensional parameter space $(\omega, K)$ by this eigenfunction is shown in \textbf{Fig.~\ref{fig:simple_kuramoto_parameter} (a)}, demonstrating the many-to-one character of the diffusion map coordinate map.

Due to the simplicity of this model it is also possible to find an effective parameter analytically. Multiplying the order parameter by $e^{-i \theta_{i}}$, taking the imaginary part, and substituting the result into the model equations yields
\begin{equation}
\frac{d\theta_i}{dt}=\omega_i+(RK_i)\sin(\psi-\theta_i),
\end{equation}
which under steady state conditions yields an effective parameter
\begin{equation}
    \theta_{i, \infty} = \psi + \arcsin\left(\frac{\omega_i}{RK_i}\right).
    \label{eq:closed_form_eff}
\end{equation}
As a validation of our data-driven approach, we compare our data-driven parameter $\phi_{1}$ to the analytical one \eqref{eq:closed_form_eff}. In \textbf{Fig.~\ref{fig:simple_kuramoto_parameter}} we show the parameter space colored by (a) our data-driven parameter $\phi_{1}$, (b) the analytical parameter, and (c) the steady state phases. As the figure illustrates, the colorings are similar suggesting that our data-driven parameter is indeed an equivalent effective parameter for this model.

This is confirmed by the plot in \textbf{Fig.~\ref{fig:simple_kuramoto_parameter} (d)}, which clearly illustrates the invertible relationship between the data-driven parameter and the analytically obtainable parameter combination. Thus, our data-driven approach is able to discover an equivalent effective parameter for this model. Combinations of the original parameters that yield the same steady state synchronized phase can be found as level sets of the eigenfunction $\phi_1$ in $(\omega, K)$ space.
\subsection*{A three-parameter example: the Kuramoto Model with Firing}
We now consider a modification to the Kuramoto model in which we additionally incorporate a ``firing term" with coefficient $\alpha_{i}$ to the coupling strength $K_{i}$, and frequency $\omega_{i}$ of each oscillator. This model was originally introduced to model excitable behavior among coupled oscillators. If $|\omega_{i}/\alpha_{i}|<1$ and $K_{i}=0$, each oscillator exhibits two steady states, one stable and one unstable. A small perturbation of the stationary, stable solution that exceeds the unstable steady state induces a firing of the oscillator, which appears as a large deviation in phase before a return to the stable state \citep{tessone2007theory,tessone2008global}. The model equations for this variation are provided below.
\begin{equation}
\frac{d\theta_i}{dt}=\omega_i+ \alpha_i\sin(\theta_i)+
\frac{K_i}{N} \sum_{j=1}^N\sin(\theta_j-\theta_i)
\label{eq:firing_kuramoto}
\end{equation}
Similar to the typical Kuramoto model, one can express the degree of phase synchronization among the oscillators with the Kuramoto order parameter,
\begin{equation}
Re^{i\psi}=\frac{1}{N}\sum_{j=1}^N e^{i\theta_j}.
\end{equation}
Transforming the model equations in the same way as those of the Kuramoto model with heterogeneous coupling coefficients \eqref{eq:kuramoto_hetero_coeffs} studied earlier results in
\begin{equation}
\frac{d\theta_i}{dt}=\omega_i+ \alpha_i\sin(\theta_i)+(RK_i)\sin(\psi-\theta_i),
\label{eq:firing_kuramoto_simplified}
\end{equation}
which under steady state conditions yields
\begin{equation}
\frac{2RK_i}{\alpha_i}\sin(\psi)\sin^2\left(\frac{\theta_i}{2}\right)+
\left(\frac{RK_i\cos(\psi)}{\alpha_i}-1\right)\sin(\theta_i)
=\frac{RK_i}{\alpha_i}\sin(\psi)+\frac{\omega_i}{\alpha_i}.
\end{equation}
Now considering a rotating reference frame, in which $\psi$ is constant, we set $\psi=0$ for convenience yielding
\begin{equation}
\left(\frac{RK_i}{\alpha_i}-1\right)\sin(\theta_i)=\frac{\omega_i}{\alpha_i}.
\end{equation}
By the above manipulations, it is now clear that the steady state phases $\theta_{i,\infty}$ are a function of $K_{i}/\alpha_{i}$ and $\omega_{i}/\alpha_{i}$, meaning that this system can be analytically described by two combinations of the original parameters. We now employ our data-driven approach to discover the effective parameter(s).

We begin by simulating this model using Scipy's vode integrator with $N=1500$ oscillators, $\alpha_i$ uniformly randomly sampled in $[-2,2]$, $K_{i}$ uniformly randomly sampled in $[10, 100]$, and $\omega_{i}$ uniformly randomly sampled in $[-\pi, \pi]$. We select these parameter values to ensure complete synchronization of the oscillators and hence a steady state in a rotating frame. As with the Kuramoto model with heterogeneous coupling coefficients, we transform the phases into a rotating frame and apply the DMAPs algorithm with an output-only informed kernel to the complex transformed steady state phases. We select diffusion map parameters of $\alpha=1$, and $\epsilon=0.9$ and find that a single diffusion map coordinate $\phi_{1}$ is identified by the local linear regression method, giving rise to a single effective parameter.

\begin{figure}[h]
    \centering
    \includegraphics[width=0.75\textwidth]{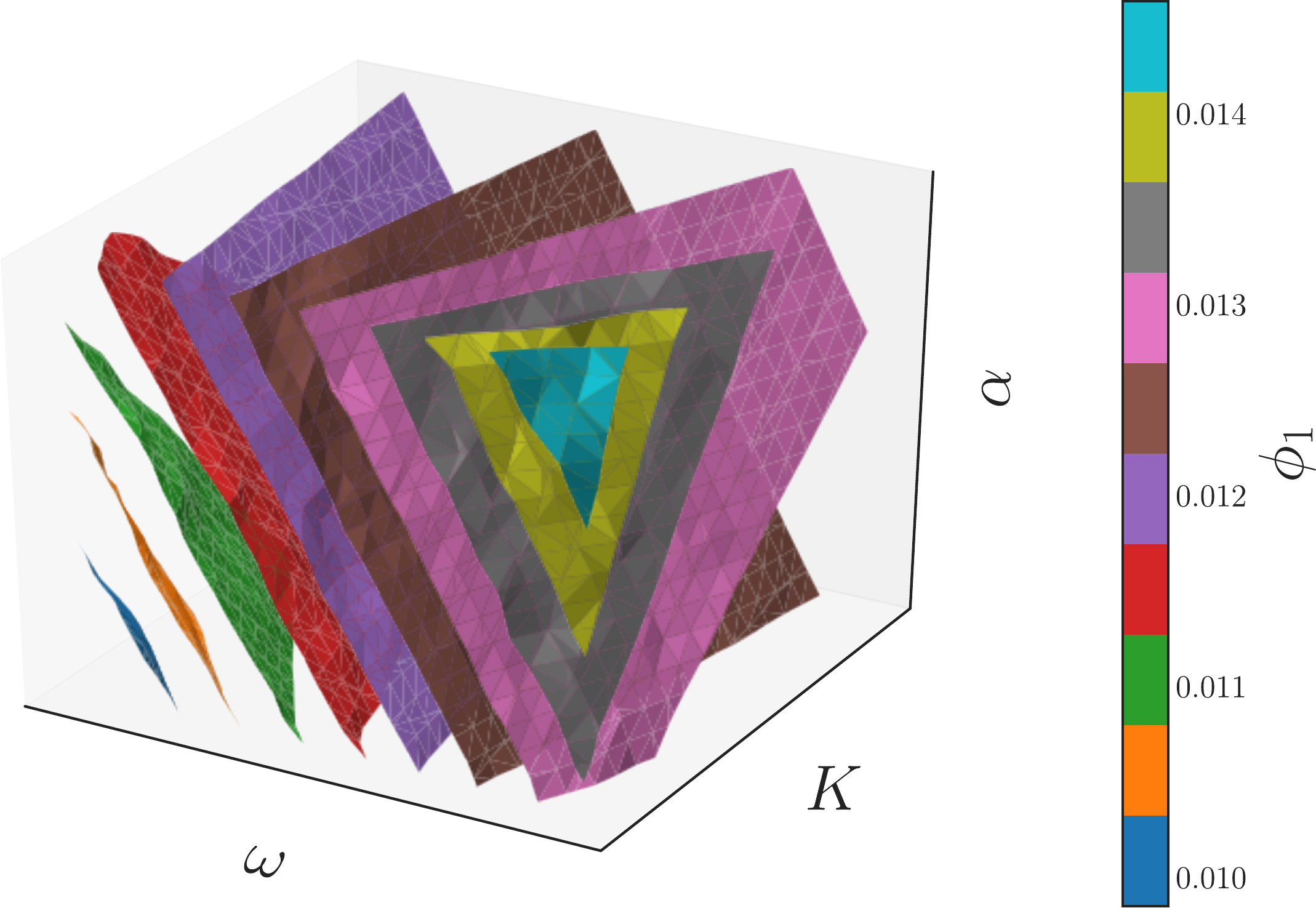}
    \caption{The level surfaces of the significant eigenfunction $\phi_{1}$ of the output-only informed diffusion map kernel in the 3D parameter space of the Kuramoto model with firing. These level surfaces were found with the marching cubes algorithm \citep{lorensen1987marching}.}
    \label{fig:2d_surf_param_3D}
\end{figure}

This parameter is a nonlinear combination of all three original parameters $(\omega, K,\alpha)$, as we illustrate in \textbf{Fig.~\ref{fig:2d_surf_param_3D}}, which shows the three-dimensional parameter space colored by the level sets of the single significant diffusion map coordinate.
Thus, this model admits a reduction of the three original parameters to a single effective parameter $\phi_{1}$, which is itself a combination of the three original system parameters $(\omega, K, \alpha)$. The relationship between the system parameters and $\phi_1$ can be subsequently explored, if desired, through standard regression techniques.
\subsection*{A more complicated example: the Kuramoto Model with Chung-Lu Coupling}
Here, we showcase our data-driven parameter discovery process for systems with more complicated couplings. Instead of the all-to-all coupled model considered by Kuramoto \eqref{eq:simple_kuramoto}, we consider a general Kuramoto model \eqref{eq:general_kuramoto} with Chung-Lu type coupling between oscillators \citep{chung2002connected}. The connection probability between oscillators for a Chung-Lu network is given by
\begin{equation}
P_{ij}=P_{ji}=\min\left(\frac{w_{i}w_{j}}{\sum_{k}w_{k}}, 1\right),
\label{eq:Chung_Lu_Prob}
\end{equation}
where $w_{i}$ is a sequence of weights defined by
\begin{equation}
w_{i}=N_p\left(1-q(i-1)/N\right)^r \quad  i=1, 2, \dots, N,
\label{eq:Chung_Lu_Weight}
\end{equation}
for network parameters $p$, $q$, and $r$. Multiple Chung-Lu networks can be generated from the same parameter values, with a specific network corresponding to generating an adjacency matrix from the connection probabilities $P_{ij}$ defined by the network parameters through the sequence of weights $w_{i}$.

\begin{figure}[h]
\begin{tabular}{ccc}
\includegraphics[width=0.3\textwidth]{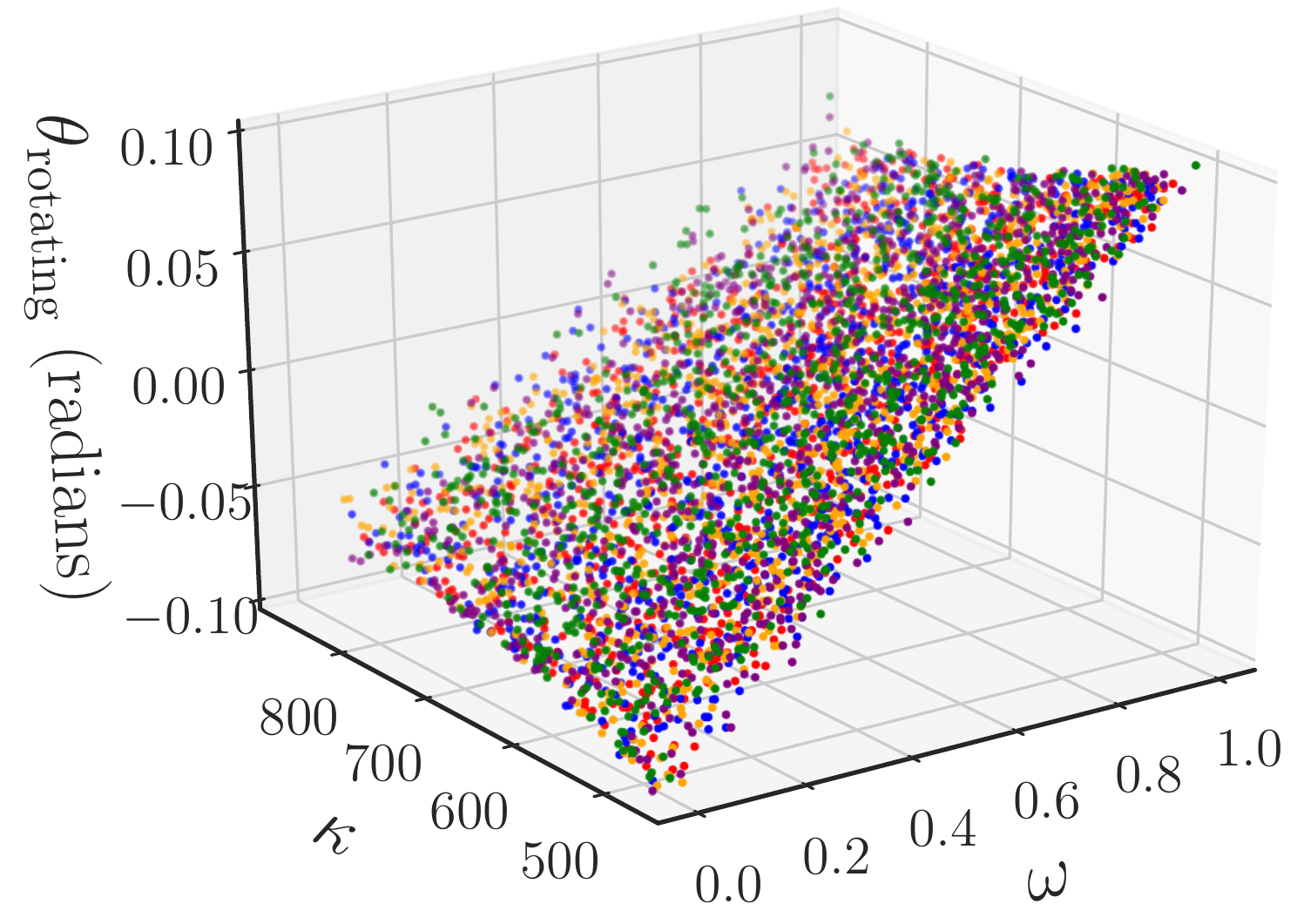} &
\includegraphics[width=0.3\textwidth]{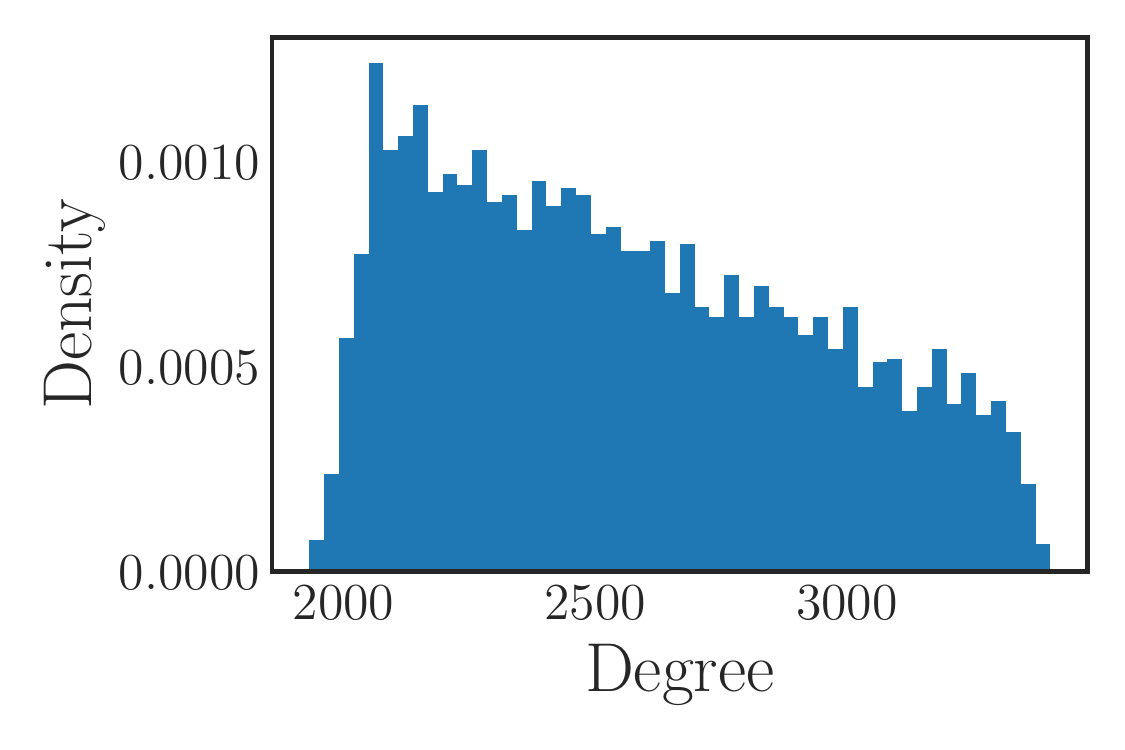} &
\includegraphics[width=0.33\textwidth]{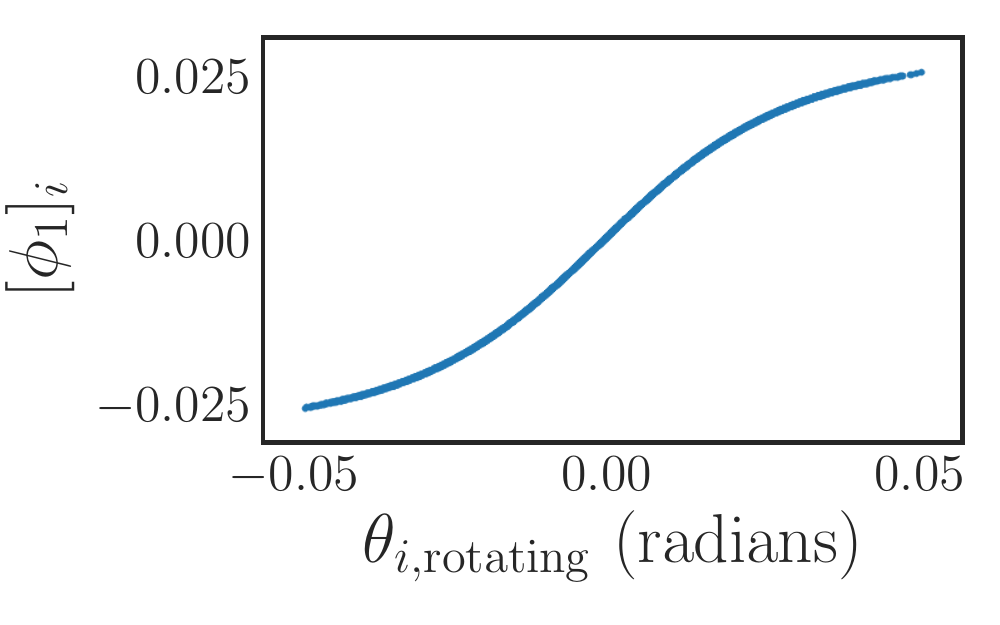} \\
(a) & (b) & (c)
\end{tabular}
\caption{(a) Invariance of the steady state phases of the Kuramoto model to different Chung-Lu networks. The dependence of the steady state phases on the model heterogeneities $(\omega, \kappa)$ for five different Chung-Lu networks generated from the same parameters is depicted, with each network corresponding to a different color. The overlapping of the differently colored plots illustrates the weak dependence of the steady state phases on the specific Chung-Lu network realization.
(b) The degree distribution of a sampling of a Chung-Lu network of $N=4000$ oscillators with parameters $p=0.5$, $q=0.9$, and $r=0.5$.
(c) The significant eigenfunction of the diffusion map as determined by the local linear regression method plotted against the steady state phases of the oscillators in a rotating frame $\theta_{i, \mathrm{rotating}}$. There is a one-to-one mapping between the two, indicating that the steady state phases depend on the single diffusion map coordinate uniquely.}
\label{fig:degree_dis_chun}
\end{figure}

One of the special properties of the Kuramoto system with a Chung-Lu coupling is that when a steady state exists in a rotating reference frame, the steady state phases of the oscillators $\theta_{i}$ lie on an invariant manifold \citep{bertalan2017coarse}, which can be parametrized by two heterogeneities: the frequency of the oscillators $\omega_{i}$, and a network property called the degree $\kappa_{i}$, as depicted in \textbf{Fig.~\ref{fig:degree_dis_chun} (a)}. This property was observed to hold for many Chung-Lu networks generated from a given set of parameters. We now show how the effect of both of these heterogeneities can be succinctly expressed by a single ``effective'' parameter.

Throughout the remainder of this section we consider a collection of 4000 oscillators ($N=4000$) with uniformly randomly distributed frequencies $\omega_{i}$ over the range $[0, 1]$ that are coupled together in a Chung-Lu network with a coupling constant of $K=20$ (multiplying the adjacency matrix), and network parameters of $p=0.5$, $q=0.9$, and $r=0.5$. \textbf{Fig.~\ref{fig:degree_dis_chun} (b)} shows the degree distribution of a sampling of a Chung-Lu network with these parameter values. We are careful to select our coupling constant large enough to ensure complete synchronization for these parameter values, and hence a steady state in a rotating frame that can be described in terms of the heterogeneities $\omega$, and $\kappa$.
After our parameter selection, we integrate the oscillators in time with SciPy's Runge-Kutta integrator (\textit{solve\_ivp} with RK45) until they reach a steady state in a rotating reference frame.

Now we show that, although there are two heterogeneity parameters, ($\omega_{i}$, $\kappa_{i}$), the long term behavior of the Kuramoto oscillators with a Chung-Lu network is intrinsically one dimensional, and can be described by a single effective parameter, which itself is a, possibly nonlinear, combination of the original system parameters.
\begin{align}
(\omega,\kappa) & \mapsto \theta_{\infty}
\quad \text{Original analysis} \\
\phi & \mapsto \theta_{\infty}
\quad \text{Diffusion map outcome}
\end{align}
In order to find this effective parameter, we begin by applying the DMAPs algorithm to the complex transformed steady state phase data with an output-only informed kernel. As mentioned before, this means that our observations consist solely of the steady state phases of the oscillators $\theta_{i, \infty}$, and ignore the oscillator heterogeneities $(\omega_{i}$, $\kappa_{i})$. For our diffusion maps we use $\alpha=1$ for the Laplace-Beltrami operator, a kernel bandwidth parameter of $\epsilon \approx 2.3 * 10^{-2}$, and the Gaussian kernel with the Euclidean distance
\begin{equation}
K_{ij} = \exp\left( - \frac {||x_i-x_j||_2^2}{\epsilon^2}\right)\quad i,j=1,\dots,4000,
\end{equation}
where $\Vert \cdot \Vert_2$ is the Euclidean norm in the complex plane of 4000 oscillator phases, i.e.~$x_j = e^{i\theta_j(t_{\infty})}$.

Next, we use the local linear regression method to determine that there is a single significant eigenfunction, $\phi_{1}$. \textbf{Fig.~\ref{fig:degree_dis_chun} (c)} shows that this eigenfuction is one-to-one with the steady state phases of the oscillators in a rotating frame, and thus provides an equivalent description of the steady state behavior of this system. Therefore, the synchronized phases of this system can be accurately described by the single effective parameter $\phi_{1}$, as claimed.

\begin{figure}[h]
    \centering
    \includegraphics[width=1.0\textwidth]{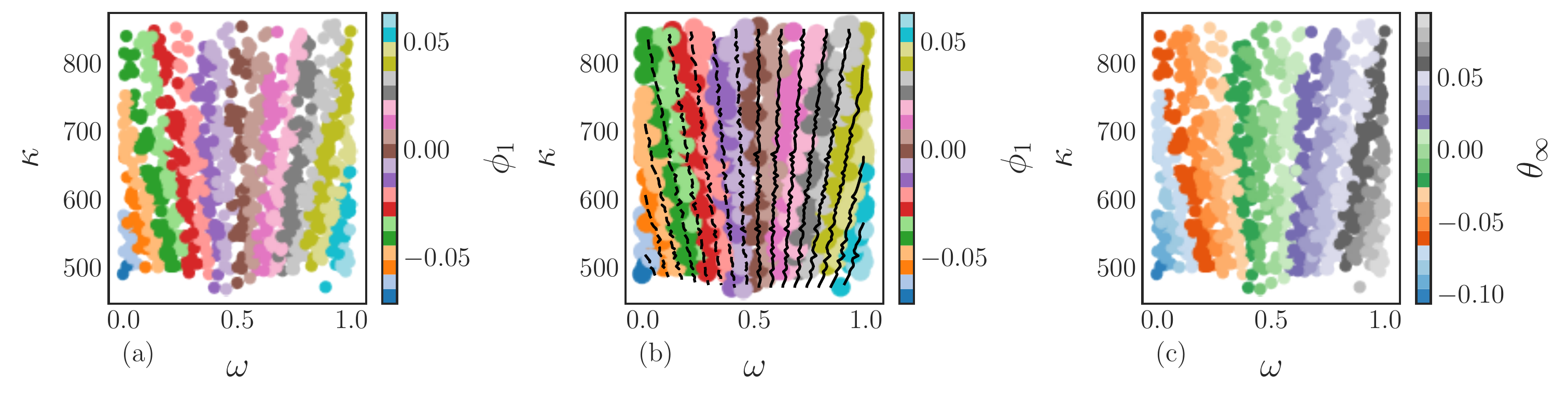}
    \caption{The application of the diffusion map algorithm to the Chung-Lu coupled Kuramoto model with an output-only-informed kernel yields a single significant diffusion map coordinate, $\phi_{1}$.
    (a) Coloring the parameter space ($\omega, \kappa$) with this new coordinate reveals the relationship between the original parameters and the ``effective'' diffusion map parameter.
    (b) Level sets of the significant diffusion map coordinate $\phi_{1}$ in the original parameter space ($\omega, \kappa$). These level sets were found by means of the marching squares algorithm \citep{maple2003geometric}. A functional form of the mapping between parameters and the diffusion map coordinate could be found with typical regression techniques or by using machine learning techniques like neural networks.
    (c) Coloring the parameter space with the steady state phases produces a coloring similar to the one in (a) as there is a one-to-one map between $\theta$ and $\phi_{1}$.}
\label{fig:preimage_parmeters}
\end{figure}

The relationship between the original parameters and the DMAPs parameter is illustrated by \textbf{Fig.~\ref{fig:preimage_parmeters} (a)}, which depicts a coloring of the original parameter space with $\phi_{1}$. In this figure one can observe that there are multiple combinations of the original parameters that correspond to the same value of $\phi_{1}$ and hence the same steady state phase. The key observation is that the level sets of $\phi_{1}$ provide the mapping between the original parameters and the new effective parameter. These level sets are depicted in \textbf{Fig.~\ref{fig:preimage_parmeters} (b)}, and can be found with established techniques, such as the marching squares algorithm \citep{maple2003geometric}. Thus, by using the DMAPs parameter $\phi_{1}$ we can express the steady state phases in terms of a single combination of the original parameters. If necessary/useful, we can try to express this new effective parameter as a function of the original system parameters through standard regression techniques, or even possibly through neural networks.

To summarize our approach, in each of three examples above we used DMAPs combined with the local linear regression method to determine the intrinsic dimensionality of the output space. The significant eigenfunctions that we obtained from this process provided new coordinates for the output space and can be considered as the ``effective'' parameters of the system. If there are fewer significant DMAPs eigenfunctions than original parameters, then this change of variables also provides a reduction in total necessary parameters. 
\section*{Discussion and Future Work}
\label{sec:Discussion}
Throughout this paper we have presented a data-driven methodology for discovering coarse variables, learning their dynamic evolution laws, and identifying sets of effective parameters. In each case, we used either an example or a series of examples to demonstrate the efficacy of our techniques compared to the established analytical technique and, in each case, the results of our data-driven approach were in close agreement with the established methodology.

Nevertheless, it is important to consider the interpretability (the ``X'' in XAI, explainable artificial intelligence) of the data-driven coarse variables discovered in our work. Even when performing model reduction with linear data-driven techniques, like Principal Component Analysis, it is difficult to ascribe a physical meaning to linear combinations of meaningful system variables (what does a linear combination of, say, a firing rate and an ion concentration ``mean''?). The conundrum is resolved by looking for physically meaningful quantities that, on the data, are one-to-one with the discovered data-driven descriptors, and are therefore equally good at parametrizing the observations. One hypothesizes a set of meaningful descriptors and then checks that the Jacobian of the transformation from data-driven to meaningful descriptors never becomes singular {\em on the data}, see \citep{frewen2011coarse, sonday2009coarse, rajendran2016data, kattis2016modeling, meila2018regression} for further discussion.

With that said, we believe that our techniques offer an approach that is both general and systematic, and we intend to apply it to a variety of coupled oscillator systems. One such problem that we are currently investigating is the possible existence, and data-driven identification, of partial differential equation (PDE) descriptions of coupled oscillator systems. Such an alternative  coarse-grained description would confer the typical benefits associated with model reduction, such as accelerated simulation and analysis; it will also present unique challenges: for instance, the selection of appropriate boundary conditions for such a data-driven PDE model of coupled oscillators. 

As we remarked in our discussion of the integrator-based neural network architecture, there is an extension of the ODE neural network approach that allows one to learn PDEs discretized by the method of lines approach \citep{gonzalez1998identificationPDE}. We plan to leverage this capability to discover PDE descriptions of coupled oscillator systems, such as the simple Kuramoto model in the continuum limit as well as of networks of Hodgkin-Huxley oscillators.

\begin{figure}[h]
    \centering
    \begin{tabular}{c}
    \includegraphics[width=0.8\textwidth]{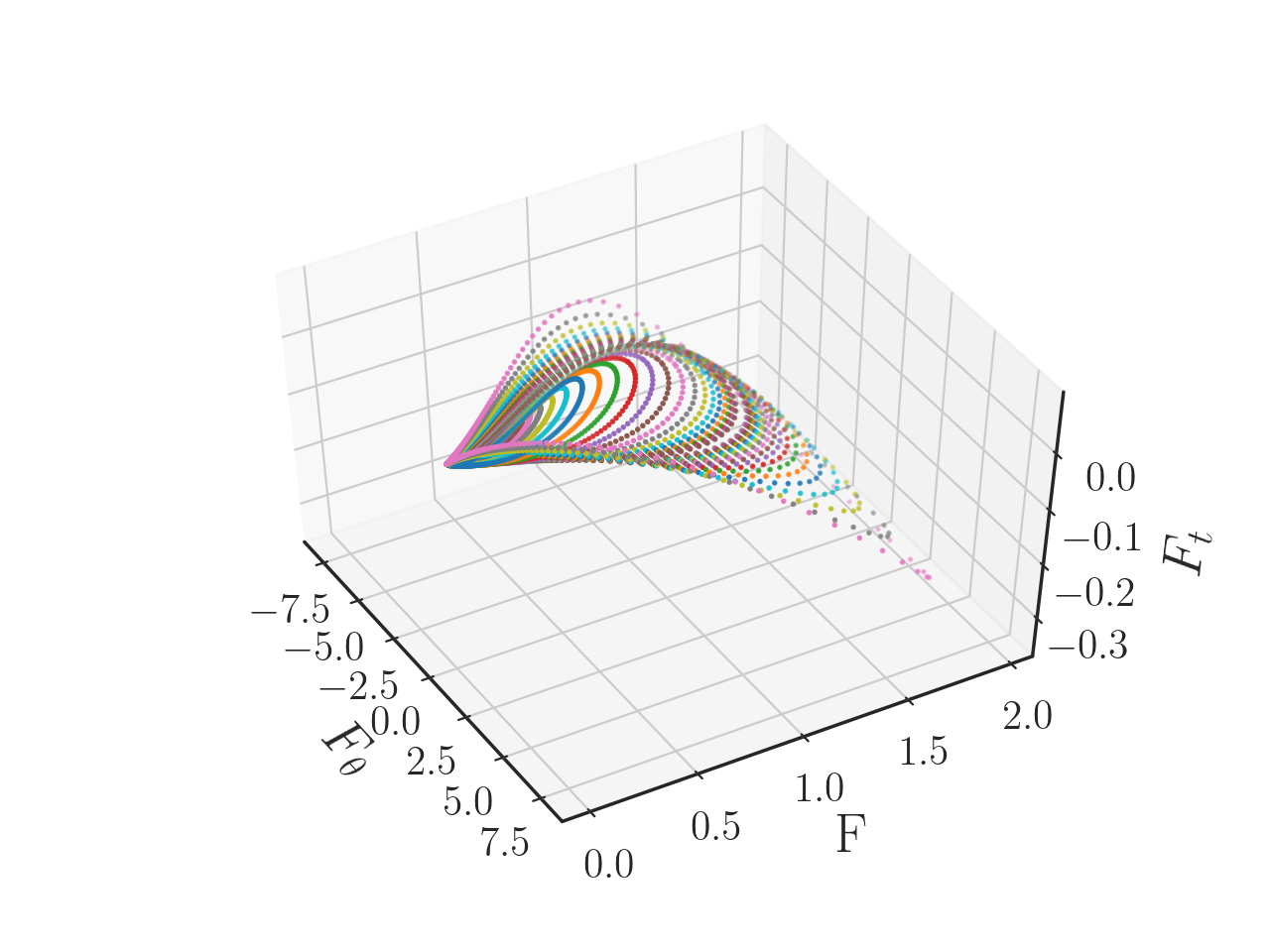} \\
    (a) \\
    \includegraphics[width=0.7\textwidth]{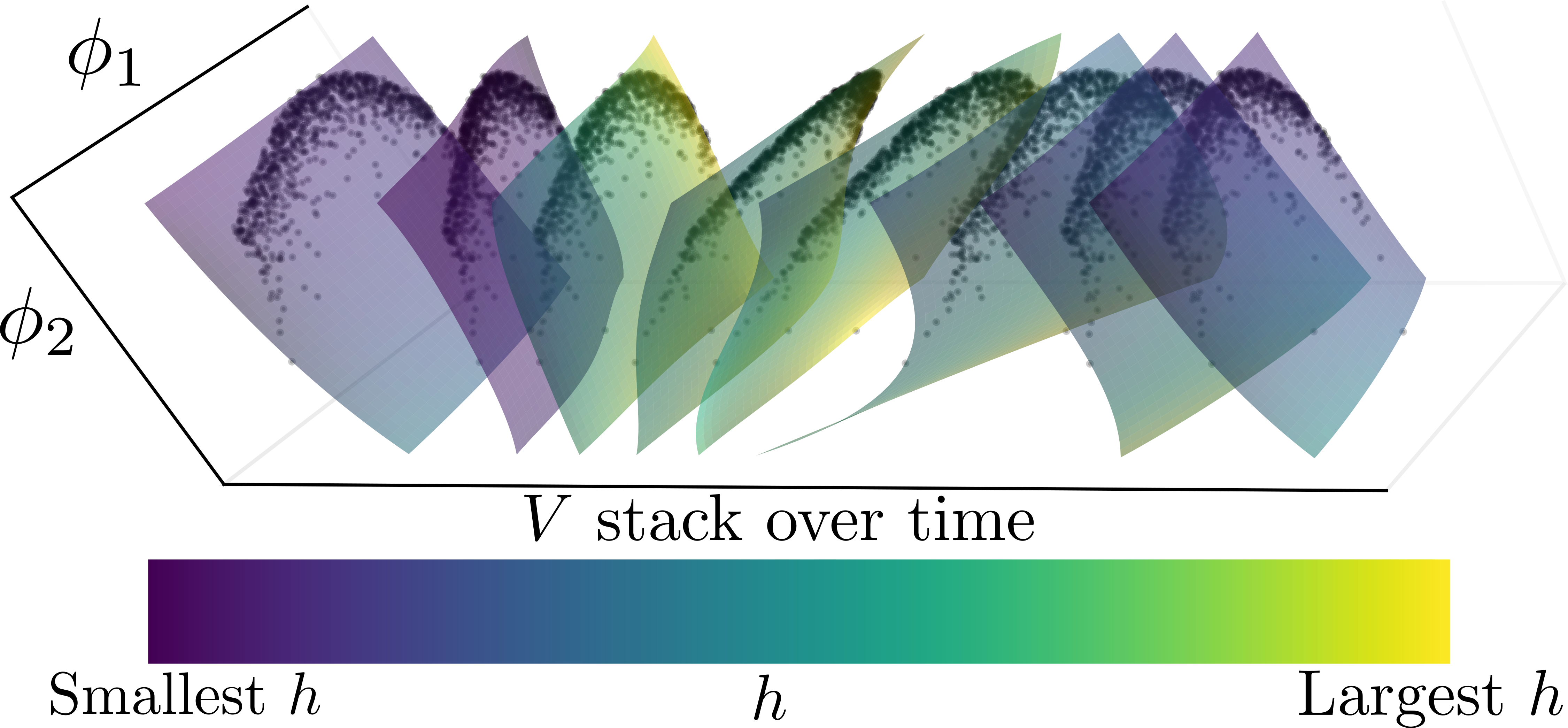} \\
    (b)
    \end{tabular}
    \caption{(a) Plot of $\frac{\partial F}{\partial t}$ versus $\frac{\partial F}{\partial \theta}$ and $F$ for a variety of $\theta$ $\in [0, 2 \pi]$ and initial $R(t) \in [0, 0.85]$ values. Each color corresponds to a different value of $R(t)$, while traversing a curve of a given color corresponds to $\theta$ running from 0 to 2$\pi$. Note that the curves appear to trace out a surface, suggesting that there is an underlying functional relationship.
    (b) Surfaces of $V$ as a function of the two discovered diffusion map coordinates, $\phi_{1}$ and $\phi_{2}$, colored by $h$ and stacked in time. This stack represents approximately one period of the limit cycle. The smoothly varying behavior of these surfaces is suggestive of a PDE for $V$ in $\phi_{1}$, $\phi_{2}$, and time. The black dots are the actual oscillators which were used to produce the surfaces in $\phi_{1}$ and $\phi_{2}$ by means of a polynomial chaos expansion.}
    \label{fig:kuramoto_PDE_1}
\end{figure}

Utilizing the properties of the continuous form of the Kuramoto model, one can express the time- and phase- dependent oscillator density $F(\theta, t)$ as an integral of the conditional oscillator density with respect to the frequency.
\begin{equation}
    F(\theta, t)=\int_{-\infty}^{\infty} \rho(\theta|\omega;t)g(\omega)d\omega
\end{equation}
As we pointed out earlier, Ott and Antonsen discovered an invariant attracting manifold for the simple Kuramoto model in the continuum limit with Cauchy distributed frequencies \citep{ott2008low}. It has been shown that on this manifold, the oscillator density $F$ satisfies the following equation.
\begin{equation}
    F(\theta, t)=\frac{1-R^{2}}{2\pi(1-2Rcos(\psi-\theta)+R^{2})}
\end{equation}
Away from this attracting invariant manifold, the full oscillator density $\rho(\theta, \omega, t)$ obeys the continuity equation
\begin{equation}
    \frac{\partial \rho(\theta, \omega, t)}{\partial t}+\frac{\partial ([\omega + \frac{K}{2i} (re^{-i \theta}-r^{*}e^{i \theta})]\rho(\theta, \omega, t))}{\partial \theta}=0
\end{equation}
We would like to use our neural network based approach to learn equivalent PDEs directly from oscillator density data. As an example of what this would look like for $F(\theta,t)$, one can approximate the partial derivative of $F$ with respect to time
\begin{equation}
    \frac{\partial F(\theta, t)}{\partial t}\approx \frac{F(\theta, t + \delta t)-F(\theta, t)}{\delta t}.
\end{equation}
Furthermore, one can numerically obtain the partial derivative of $F$ with respect to $\theta$. Plotting $\frac{\partial F}{\partial t}$ versus both $F$ and $F_{\theta}$ produces a loop, as illustrated in \textbf{Fig.~\ref{fig:kuramoto_PDE_1}}, where traversing a loop corresponds to varying $\theta$ from 0 to 2$\pi$. Each of the different loops depicted in \textbf{Fig.~\ref{fig:kuramoto_PDE_1}} coincides with a different initial value of $R(t)$, ranging from 0 to 0.85. Thus, it appears that along the attracting manifold, we can write
\begin{equation}
    \frac{\partial F}{\partial t}=G\Big(F,\frac{\partial F}{\partial \theta}\Big)
\end{equation}
where $G$ is an unknown function of $F$ and $\frac{\partial F}{\partial \theta}$.
The unknown function $G$ takes a form that is amenable to being learned with the PDE extension of the neural network integration procedure. In future work we intend to demonstrate this process.

Moving away from simple phase oscillators, we consider a model of Hodgkin-Huxley neural oscillators studied in \citep{choi2016dimension}, which are characterized by two state variables each, a channel state $h$ and a potential $V$ \eqref{eq:hh_neurons},
\begin{equation}
\begin{aligned}
    C\frac{dV_{i}}{dt}&=-g_{Na}m(V_{i})h_{i}(V_{i}-V_{Na})-g_{l}(V_{i}-V_{l})+I^{i}_{syn}+I^{i}_{app},
    \label{eq:hh_neurons} \\
    \frac{dh_{i}}{dt}&=\frac{h_{\infty}(V_{i})-h_{i}}{\tau(V_{i})},
\end{aligned}
\end{equation}
for $i=1,\dots,N$. Where the coupling is provided by the synaptic current $I_{syn}$ defined as
\begin{equation}
    I^{i}_{syn}=\frac{g_{syn}(V_{syn}-V_{i})}{N}\sum^{N}_{j=1}A_{ij}s(V_{j}),
\end{equation}
with adjacency matrix $A_{ij}$. The functions $\tau(V)$, $h_{\infty}(V)$, and $m(V)$ are a standard part of the Hodgkin-Huxley formalism, $s(V)$ is the synaptic communication function, and $g_{Na}$, $V_{Na}$, $g_{l}$, and $V_{l}$ are model parameters.

It has been shown that with a Chung-Lu network (\ref{eq:Chung_Lu_Prob}, \ref{eq:Chung_Lu_Weight}) these oscillators are drawn to an attractive limit cycle along which their states can be described by two parameters: their applied current $I_{app}$ and their degree, $\kappa$. These two heterogeneities can also be described by two diffusion map coordinates, $\phi_{1}$ and $\phi_{2}$ \citep{kemeth2018emergent}. Plotting the potential, $V$, for a single period of the limit cycle produces the stack of surfaces shown in \textbf{Fig.~\ref{fig:kuramoto_PDE_1} (b)}. The smoothly varying character of these surfaces is suggestive of a PDE description of these oscillators along this limit cycle. We intend to investigate the identification of such a PDE through an extension of the data-driven identification technique for coarse ODEs presented herein.
\section*{Conflict of Interest Statement}
The authors declare that the research was conducted in the absence of any commercial or financial relationships that could be construed as a potential conflict of interest.
\section*{Author Contributions}
I.G.K and C.R.L. planned the research. T.N.T, M.K. and T.B. performed the computations, and coordinated their design with I.G.K. and C.R.L. All authors contributed to writing, editing and proofreading the manuscript. 
\section*{Funding}
The work was partially supported by the DARPA PAI program, an ARO-MURI as well as the NIH through a UO1 grant. MK acknowledges funding by SNSF grant P2EZP2\_168833. 
\section*{Acknowledgments}
The authors are pleased to acknowledge fruitful discussions with Drs. F. Dietrich and J. Bello-Rivas, as well as the use of a diffusion maps/geometric harmonics computational package authored by the latter.
\section*{Data Availability Statement}
The raw data supporting the conclusions of this manuscript will be made available by the authors, without undue reservation, to any qualified researcher.

\bibliographystyle{frontiersinSCNS_ENG_HUMS} 
\bibliography{dmaposcil}




\end{document}